# Markov Decision Process Based Energy-Efficient On-Line Scheduling for Slice-Parallel Video Decoders on Multicore Systems

Nicholas Mastronarde, Karim Kanoun,
David Atienza, Pascal Frossard, and Mihaela van der Schaar

### Abstract

We consider the problem of energy-efficient on-line scheduling for slice-parallel video decoders on multicore systems. We assume that each of the processors are Dynamic Voltage Frequency Scaling (DVFS) enabled such that they can independently trade off performance for power, while taking the video decoding workload into account. In the past, scheduling and DVFS policies in multi-core systems have been formulated heuristically due to the inherent complexity of the on-line multicore scheduling problem. The key contribution of this report is that we rigorously formulate the problem as a Markov decision process (MDP), which simultaneously takes into account the on-line scheduling and per-core DVFS capabilities; the power consumption of the processor cores and caches; and the loss tolerant and dynamic nature of the video decoder's traffic. In particular, we model the video traffic using a Direct Acyclic Graph (DAG) to capture the precedence constraints among frames in a Group of Pictures (GOP) structure, while also accounting for the fact that frames have different display/decoding deadlines and non-deterministic decoding complexities.

The objective of the MDP is to minimize long-term power consumption subject to a minimum Quality of Service (QoS) constraint related to the decoder's throughput. Although MDPs notoriously suffer from the curse of dimensionality, we show that, with appropriate simplifications and approximations, the complexity of the MDP can be mitigated. We implement a slice-parallel version of H.264 on a multiprocessor ARM (MPARM) virtual platform simulator, which provides cycle-accurate and bus signal-accurate simulation for different processors. We use this platform to generate realistic video decoding traces with which we evaluate the proposed on-line scheduling algorithm in Matlab.

## 1. Introduction

High-quality video decoding imposes unprecedented performance requirements on energy-constrained mobile devices. To address the competing requirements of high performance and energy-efficiency, embedded mobile multimedia device manufactures have recently adopted MPSoC (multiprocessor system-on-chip) architectures that support Dynamic Voltage Frequency Scaling (DVFS) and Dynamic Power Management (DPM) technologies. DVFS enables dynamic adaption of each processor's frequency and voltage, and can be exploited to reduce power consumption when the maximum frequency of operation is not required to meet the deadlines of a certain set of tasks [9]. Meanwhile, DPM enables system components such as processors to be dynamically switched on and off, and can be exploited to reduce leakage power consumption when these components are not needed [17].

Despite improvements in mobile device technology, energy-efficient multicore scheduling for video decoding remains a challenging problem for several reasons. First, video decoding applications have intense and time-varying stochastic workloads, which have worst-case execution times that are significantly larger than the average case. Second, video applications have sophisticated dependency structures due to predictive coding. These dependency structures, which can be modeled as directed acyclic graphs (DAGs), not only result in different frames having different priorities, but also make it difficult to balance loads across the cores, which is important for energy efficiency [1]. Finally, video applications often have stringent delay constraints, but are considered soft real-time applications [21]. In other words, video frames should meet their deadlines, but when they do not, the application quality (e.g. decoded video frame rate) is reduced.

During the last decade, many energy-efficient multicore scheduling algorithms that exploit DVFS and/or DPM have been proposed, e.g. [3][4][5][7][8][10][11]. In Table 1, we classify these representative solutions based on their utilized optimization horizons, application models, complexity models, scheduling granularities, supported power management schemes, and considered sources of energy/power. We note that [3] and [4] are specifically designed for video decoding; [7],[8], and [11] use video decoding as an illustrative application; and [5] and [10] are not designed for video decoding.

The Largest Task First with Dynamic Power Management (LTF-DPM) algorithm in [4] assumes that frame decoding deadlines are equally spaced in time (e.g. 33 ms apart for 30 frame per second video), and therefore does not support video group of pictures (GOP) structures with B frames; moreover, LTF-DPM will typically have looser deadline constraints than our proposed algorithm because it assigns groups of frames a common "weak" deadline.

The Scheduling2D and Stochastic Scheduling2D algorithms in [7] and [8], respectively, can be applied to video decoding applications, but both consider a periodic directed acyclic graph (DAG) application model that requires a "source" and "sink" node in each period, making the algorithms incompatible with GOP structures where the last B frame in a GOP depends on the I frame in the next GOP (e.g. an IBPB GOP).

The Variation Aware Time Budgeting (Var-TB) algorithm in [11] uses a DAG task model and allows for arbitrary complexity distributions; however, the author's propose using a functional partitioning algorithm for parallelizing the video decoder (e.g. pipelining decoder subfunctions such as inverse DCT and motion compensation on different cores). Functional partitioning is known to be suboptimal because moving data between cores requires a lot of memory bandwidth [19]. It is shown in [19] that parallelization approaches based on data partitioning (e.g. mapping different frames, slices, or macroblocks to different processors) are superior to functional partitioning approaches [19].

The Global Earliest Deadline First Online DVFS (GEDF-OLDVFS) algorithm in [10] is inappropriate for predictively coded video applications because it assumes that tasks are independent. Finally, the so-called SpringS algorithm in [5] uses a task-level software pipelining algorithm called RDAG [6] to transform a periodic dependent task graph (expressed as a DAG) into a set of tasks that can be pipelined on parallel processors. Unfortunately, if this technique is applied to video decoding applications, it will require retiming delays proportional to the GOP size, which may be arbitrarily large.

All prior research outlined in Table 1 takes into account processing energy, but does not take into account the power consumption of different cache levels in the memory hierarchy. Since multimedia applications are data-access dominated [12], read and write accesses to the memory cache contribute significantly to the overall energy consumption.

**Table 1. A qualitative comparison to representative related work.**

| Method | Optimization Horizon | Application model | Complexity model | Scheduling granularity | Supported power management | Considered sources of energy/power |
|--------|----------------------|-------------------|------------------|------------------------|----------------------------|------------------------------------|
| OPT-MEMS [3] | Myopic | Periodic real-time tasks | Worst-case execution time | Fluid | DPM and coordinated DVFS | Processing |
| LTF-DPM [4] | Finite | Periodic real-time tasks | Predicted computational complexity | Video slice | DPM and independent DVFS | Processing |
| Scheduling2D [7] SScheduling2D [8] | Finite | Periodic DAG with source and sink | Worst-case execution time (Average-case execution time) | DAG node (e.g. video frame or slice) | DPM and independent DVFS | Processing |
| VAR-TB [11] | Finite | DAG | Arbitrary distribution | DAG node (decoding subfunction) | Independent DVFS | Processing |
| GEDF-OLDVFS [10] | Myopic | General independent real-time tasks | Worst-case execution time | Task | DPM and independent DVFS | Processing |
| SpringS + RDAG [5] | Myopic | Periodic DAG with source and sink | Known computational complexity | DAG node | DPM and independent DVFS | Processing |
| Proposed | Infinite | Periodic Markov chain | Exponentially distributed | Video slice | Independent DVFS | Processing + data access |
| Proposed w/ coordinated DVS | Infinite | Periodic Markov chain | Exponentially distributed | Video slice | Coordinated DVFS | Processing + data access |

In summary, although many important advancements have been made, there is still no rigorous multicore scheduling solution that simultaneously considers per-core DVFS capabilities; dynamic processor assignment; the separate power consumption of the processor cores and caches; and loss-tolerant tasks with different complexity distributions, DAG dependency structures (i.e. precedence constraints), and stringent, but soft real-time, constraints. The contributions of this report are as follows:

- We rigorously formulate the multi-core scheduling problem using a Markov decision process (MDP) that considers the abovementioned properties of the multi-core system and video decoding application. The MDP enables the system to optimally trade-off long-term power and performance, where the performance is measured in terms of a Quality of Service (QoS) metric that is related to the decoder's throughput.

- The MDP solution requires complexity that exponentially increases with both the number of processors and the number of frames in a short look-ahead window. To mitigate this complexity, we propose a novel two-level scheduler. The first-level scheduler determines scheduling and DVFS policies for each frame using frame-level MDPs, which account for the coupling between the optimal policies of parent frames and their children's optimal policies. The first-level acts in discrete time. The second-level scheduler decides the final frame-to-processor and frequency-to-processor mappings at run-time, ensuring that certain system constraints are satisfied. The second-level also performs slack reclamation [8][10][11] to avoid wasting resources when tasks finish before the first-level scheduler's time quantum is up.

- We validate the proposed algorithm in Matlab using accurate video decoder trace statistics generated from a parallelized H.264 decoder that we implemented on a cycle-accurate MPARM simulator [15].

The remainder of the report is organized as follows. We introduce the system and application models in Section 2 and formulate the on-line multi-core scheduling problem as an MDP. In Section 3, we propose a lower complexity solution by approximating the original MDP problem with a two-level scheduler. In Section 4, we present our experimental results. We conclude in Section 5.

## 2. PROBLEM FORMULATION

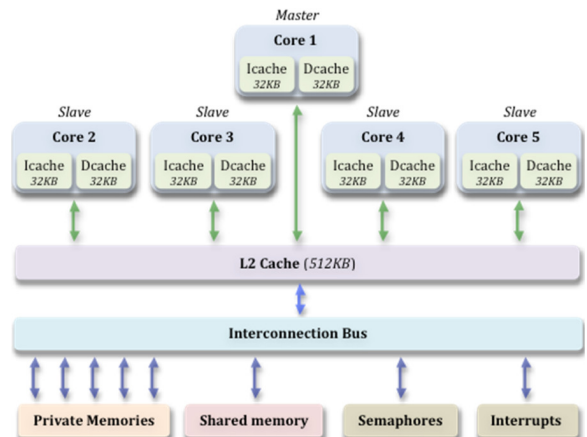

**Fig. 1. Hardware configuration of our MPARM-based virtual platform.**

We consider the problem of energy-efficient slice-parallel video decoding in a time slotted multicore system, where time is divided into slots of (equal) duration $\Delta t$ seconds indexed by $t \in \mathbb{N}$. We assume that there are $M$ slave processors, which we index by $j \in \{1, \ldots, M\}$, and one master processor as illustrated in Fig. 1. Our problem formulation focuses on scheduling slice decoding tasks on the slave cores. We discuss the master core in more detail in Section 4.

In Section 2.1, we describe seven important video data attributes. In Section 2.2, we propose a sophisticated Markovian traffic model for characterizing video decoding workloads. Importantly, the proposed traffic model accounts for the video data attributes introduced in Section 2.1. In Section 2.3, we use the traffic model to reveal several opportunities for parallel execution of slice decoding tasks. In Sections 2.4, 2.5, and 2.6 we describe the scheduling and frequency actions, the evolution of the video traffic/workload, and the power and Quality of Service (QoS) metrics used in our optimization. In subsection 2.7, we formulate the multicore scheduling problem as a Markov decision process (MDP).

## 2.1. Video data attributes

We model the encoded video bitstream as a sequence of compressed data units with different decoding and display deadlines, source-coding dependencies, priorities, and decoding complexity distributions. In this report, we assume that a data unit corresponds to one video slice, which is a subset of a video frame that can be decoded independently of other slices within the same frame.[1]

We assume that the video is encoded using a fixed, periodic, GOP structure that contains $K$ frames and lasts a period of $T$ time slots of duration $\Delta t$. The set of frames within GOP $g \in \mathbb{N}$ is denoted by $\mathcal{V}^g \triangleq \left\{ v_1^g, v_2^g, \ldots, v_K^g \right\}$ and the set of all frames is denoted by $\mathcal{V} \triangleq \bigcup_{g \in \mathbb{N}} \mathcal{V}^g$. Each frame $v_k^g$ is characterized by seven attributes:

1. *Type*: Frame $v_k^g$ is an I, P, or B frame.[2] We denote the operator extracting the frame type by $\mathrm{type}\left( v_k^g \right)$.

2. *Number of slices*: Frame $v_k^g$ is composed of $l^{v_k^g} \in \left\{ 1, \ldots, l^{\max} \right\}$ slices, where $l^{v_k^g}$ is assumed to be fixed[3] and $l^{\max}$ is the maximum number of slices allowed in any single video frame. The number of slices $l^{v_k^g}$ is determined by the encoder.

3. *Decoding complexity*: Slices belonging to frame $v_k^g$ have decoding complexity $w^{v_k^g}$ cycles. We assume that $w^{v_k^g}$ is an exponentially distributed i.i.d. random variable conditioned on the frame type with expectation $\mathbf{E}\left[ w^{v_k^g} \right] = \beta^{\mathrm{type}\left( v_k^g \right)}$. The assumption of exponentially distributed complexity is inaccurate; however, it is necessary to make the MDP problem formulation tractable. We briefly discuss why we make this assumption in Section 2.4 and its consequences in Section 3.3, and provide further details in Appendix A.

---

[1] Because slices within a frame are encoded without exploiting correlations among neighboring slices, there is a trade-off between video rate-distortion performance during encoding (which is better for coarser grained slices) and potential parallelization gains during decoding (which are higher for finer grained slices). This trade-off has been thoroughly discussed in prior work [13]. The focus of this report is on optimally scheduling slices at the decoder side given a bitstream that has already been encoded with slices.

[2] In a typical hybrid video coder like H.264/AVC or MPEG-2, I, P, and B indicate the type of motion prediction used to exploit temporal correlations between video frames. I-frames are compressed independently of the other frames, P-frames are predicted from previous frames, and B-frames are predicted from previous and future frames.

[3] For simplicity of exposition, we assume that the bitstream is pre-encoded and that it was encoded using a fixed number of slices per frame. However, our framework can be adapted to account for an encoder that uses a variable number of slices per frame (e.g. by generating slices of approximately equal computational complexity [13] or equal size in bits). If the video has been pre-encoded, then we can assume that $l^{v_k^g}$ is known. Alternatively, if the encoded bitstream is generated in real-time (as in a video conferencing application), then, at the decoder, we can model $l^{v_k^g}$ as an i.i.d. random variable conditioned on the frame type (i.e. $\mathrm{type}\left( v_k^g \right)$) or position of the frame in the GOP (i.e. $k$).

4. *Arrival time*: $t^{v_k^g}$ denotes the earliest time slot in which $v_k^g$ can be decoded (i.e., its arrival time at the scheduler).

5. *Display deadline*: $d^{v_k^g,\mathrm{disp}}$ denotes the final time slot in which $v_k^g$ must be decoded so that it can be displayed.

6. *Decoding deadline*: $d^{v_k^g,\mathrm{dec}}$ denotes the final time slot in which $v_k^g$ must be decoded so that frames that depend on it can be decoded before their display deadline. Note that $d^{v_k^g,\mathrm{dec}} \le d^{v_k^g,\mathrm{disp}}$.

7. *Dependency*: The frames must be decoded in decoding order, which is dictated by the dependencies introduced by predictive coding (e.g., motion-compensation). In general, the dependencies among frames can be described by a directed acyclic graph (DAG), denoted by $DAG \triangleq \langle \mathcal{V}, \mathcal{E} \rangle$, with the nodes in $\mathcal{V}$ representing frames and the edges in $\mathcal{E}$ representing the dependencies among frames. We use the notation $v_{k'}^{g'} \prec v_k^g$ to indicate that frame $v_k^g$ depends on frame $v_{k'}^{g'}$ (i.e., there exists a path directed from $v_{k'}^{g'}$ to $v_k^g$) and therefore $v_k^g$ cannot be decoded until $v_{k'}^{g'}$ is decoded.[4] We write $\left( v_{k'}^{g'}, v_k^g \right) \in \mathcal{E}$ if there is a directed arc emanating from frame $v_{k'}^{g'}$ and terminating at frame $v_k^g$, indicating that $v_{k'}^{g'}$ is an immediate parent of $v_k^g$.

These attributes are important because they determine which slices can be decoded, how long they will take to decode, when they need to be decoded, and what the penalty is for not decoding them on time. In the next subsection, we propose a Markovian traffic model that captures the above attributes in a structured manner, enabling us to rigorously formulate the multicore scheduling problem as an MDP.

### 2.2. Markovian Traffic Model

We define a *traffic state* $\mathcal{T}_t = \left( \mathcal{C}_t, \mathbf{x}_t, \mathbf{r}_t \right)$ to represent the video data that can potentially be decoded in time slot $t$. This traffic state comprises three components defined in the following paragraphs: the *current frame set* $\mathcal{C}_t \subset \mathcal{V}$, the *buffer state* $\mathbf{x}_t$, and the *dependency state* $\mathbf{r}_t$.

In time slot $t$, we assume that the set of frames whose deadlines are within the *scheduling time window* (STW) $\left[ t, t + W_t \right]$ can be decoded. We define current frame set as all the frames within the STW, i.e.

---

[4] Note that frames in GOP $g+1$ do not depend on frames in GOP $g$; however, frames in GOP $g$ can depend on frames in GOP $g+1$ (e.g. the last B frames in GOP $g$ may depend on the I frame in GOP $g+1$).

$\mathcal{C}_t = \left\{ v \in \mathcal{V} \mid d_v \in \left\{ t, t+1, \ldots, t+W_t \right\} \right\}$. Because the GOP structure is fixed and periodic, $\mathcal{C}_t$ is periodic with some period $T$. Frame $v$'s arrival time $t^v$, display deadline $d^{v,\mathrm{disp}}$, and decoding deadline $d^{v,\mathrm{dec}}$ are fully determined by the periodic GOP structure. Specifically, it turns out that $t^v \triangleq \min\left\{ t \mid v \in \mathcal{C}_t \right\}$, $d^{v,\mathrm{disp}} \triangleq \max\left\{ t \mid v \in \mathcal{C}_t \right\}$, and $d^{v,\mathrm{dec}} \triangleq \min\left\{ d^{u,\mathrm{disp}} \mid (v,u) \in \mathcal{E} \right\}$. In words, a frame's arrival time (respectively, display deadline) is the first (respectively, last) time slot in which it appears in the current frame set, and a frame's decoding deadline is the minimum display deadline of its children. Note that the distinction between display and decoding deadlines is important because, even if a frame's decoding deadline is missed, which renders its children undecodable, it is still possible to decode the frame before its display deadline. Fig. 2 illustrates how the current frame sets are defined for a simple IBPB GOP structure and Table 2 tabulates the decoding and display deadlines for the same GOP structure. The following example illustrates one way to define the current frame sets for the GOP structure in Fig. 2.

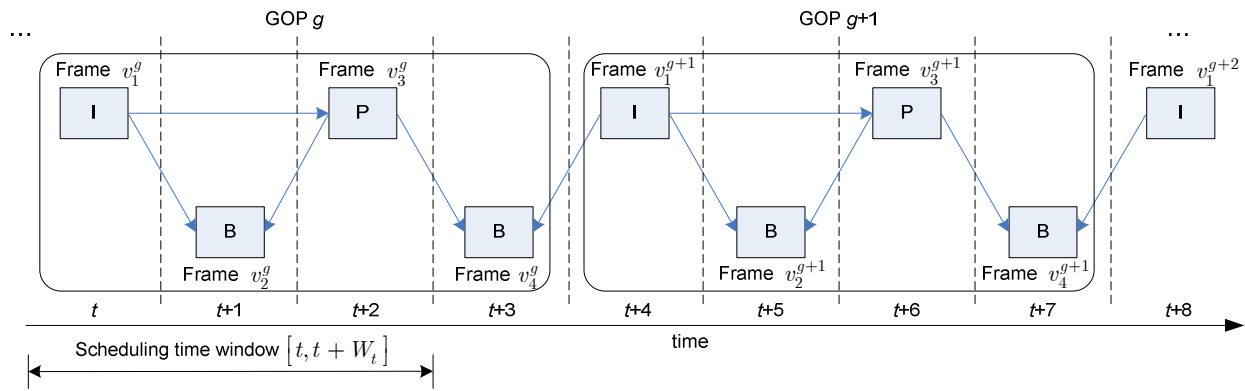

**Fig. 2. Illustrative DAG dependencies for an IBPB GOP structure that contains $K = 4$ frames and lasts a period of $T = 4$ time slots of duration $\Delta t = 1/30$ seconds.**

**Table 2. Decoding and display deadlines for the GOP structure in Fig. 2.**

|  | $v_1^g$ (I) | $v_2^g$ (B) | $v_3^g$ (P) | $v_4^g$ (B) | $v_1^{g+1}$ (I) | $v_2^{g+1}$ (B) | $v_3^{g+1}$ (P) | $v_4^{g+1}$ (B) |
|---|---|---|---|---|---|---|---|---|
| Decoding Deadline | t | t+1 | t+1 | t+3 | t+3 | t+5 | t+5 | t+7 |
| Display Deadline | t | t+1 | t+2 | t+3 | t+4 | t+5 | t+6 | t+7 |

**Example 1: Current frame sets:** Let $W_t = W_{t+2} = 2$ and $W_{t+1} = W_{t+3} = 3$. *Using the GOP structure in Fig. 2, and a time slot duration of $\Delta t = 1/30$ s, the current frame sets defined by these scheduling time windows are* $\mathcal{C}_t = \left\{ v_1^g, v_2^g, v_3^g \right\}$, $\mathcal{C}_{t+1} = \left\{ v_2^g, v_3^g, v_4^g, v_1^{g+1} \right\}$, $\mathcal{C}_{t+2} = \left\{ v_3^g, v_4^g, v_1^{g+1} \right\}$, $\mathcal{C}_{t+3} = \left\{ v_4^g, v_1^{g+1}, v_2^{g+1}, v_3^{g+1} \right\}$, *and* $\mathcal{C}_{t+4} = \left\{ v_1^{g+1}, v_2^{g+1}, v_3^{g+1} \right\}$. *Notice that the GOP structure is periodic with period $T = 4$ such that the* *current frame sets $\mathcal{C}_t$ and $\mathcal{C}_{t+T}$ contain frames in the same position of the GOP with the same underlying dependency structure.*

We define the buffer state $\mathbf{x}_t = \left( x_t^v \mid v \in \mathcal{C}_t \right)$, where $x_t^v$ denotes the number of slices of frame $v$ awaiting decoding at time $t$. By definition, $x_t^v \leq l^v$, where $l^v$ is the total number of slices belonging to frame $v$. Fig. 3 illustrates the definition of the buffer state. Finally, the dependency state $\mathbf{r}_t \triangleq \left( r_t^v \mid v \in \mathcal{C}_t \right)$ defines whether or not each frame in the current frame set is decodable in time slot $t$. In particular, $r_t^v$ is a binary variable that takes value 1 if all of frame $v$'s dependencies are satisfied, i.e. if $x_{u,t} = 0$ for all $u \prec v$, and takes value 0 otherwise. We describe how the current frame set, buffer state, and dependency state evolve from time slot to time slot in Section 2.5.

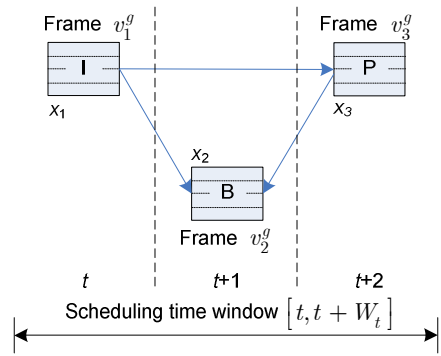

**Fig. 3. Buffer state for the current frame set** $\mathcal{C}_t = \left\{ v_1^g, v_2^g, v_3^g \right\}$. $x_k$ **denotes the number of slices belonging to frame** $v_k^g \in \mathcal{C}_t$. **The slice boundaries are shown as dotted lines in each frame.**

*2.3. Opportunities for parallelism*

Given the current frame sets illustrated in Example 1 and the GOP structure in Fig. 2, the following example identifies four opportunities for parallelism.

**Example 2: Opportunities for parallelism:**

1. *Slices from the same frame can be decoded in parallel;*

2. *Slices belonging to I and P frames at the GOP boundary (e.g. $v_3^{i,g}$ and $v_1^{i,g+1}$ in Fig. 2) can be decoded in parallel after their dependencies are satisfied;*

3. *Slices belonging to certain B and I frames (e.g. $v_2^g$ and $v_1^{g+1}$ in Fig. 2) and certain B and P frames (e.g. $v_4^g$ and $v_3^{g+1}$ in Fig. 2) can be decoded in parallel after their dependencies are satisfied.*

*4. Slices belonging to two different B frames (e.g. $v_2^{i,g}$ and $v_4^g$ in Fig. 2) can be decoded in parallel after their dependencies are satisfied.*

The goal in the proposed framework is to optimally and dynamically map slices to processors, and adapt the processors' frequencies, in order to minimize average system power consumption subject to a minimum average slice decoding rate. This optimization is made more formal in section 2.7; however, before we can formalize the optimization, we need to define the scheduling and frequency actions, model the evolution of the video traffic/workload over time, and define the power and QoS metrics used in our optimization.

*2.4. Scheduling actions and processor frequencies*

Let $y_t^{jv} \in \mathcal{Y} = \{0,1\}$ denote the number of slices belonging to frame $v$ that are scheduled on processor $j$ at time $t$. For notational convenience, we define $\mathbf{Y}_t = \left[ y_t^{jv} \right]_{jv}$, $\mathbf{y}_t^v \triangleq \left( y_t^{jv} \mid j \in \{1,...,M\} \right)^T$, and $\mathbf{y}_t^j = \left( y_t^{jv} \mid v \in \mathcal{C}_t \right)$. There are three important constraints on the scheduling actions $y_t^{jv}$ for all $j \in \{1,...,M\}$ and $v \in \mathcal{C}_t$:

- *Buffer constraint*: $\sum_{j=1}^M y_t^{jv} \leq x_t^v$. In words, the total number of scheduled slices belonging to frame $v$ cannot exceed the number of slices in frame $v$'s buffer in time slot $t$.

- *Processor constraint*: $\sum_{v \in \mathcal{C}_t} y_t^{jv} \leq 1$. In words, no more than one slice can be scheduled on processor $j$ in time slot $t$.

- *Dependency constraint*: If $r_t^v = 0$, then $\sum_{j=1}^M y_t^{jv} = 0$. In words, all of the $v$th frame's dependencies must be satisfied before slices belonging to it are scheduled to be decoded.

We assume that each processor can operate at a different frequency in each time slot to trade-off processing energy and delay. Let $\mathbf{f}_t \triangleq \left( f_t^1, f_t^2, ..., f_t^M \right) \in \mathcal{F}^M$ denote the *frequency vector*, where $f_t^j \in \mathcal{F}$ is the speed of the $j$th processor in time slot $t$ and $\mathcal{F}$ is the set of available operating frequencies. Recall from Section 2.1 that slices belonging to frame $v$ have decoding complexity $w^v$ cycles, where $w^v$ is assumed to be exponentially distributed with mean $\mathbf{E}\left[ w^v \right] = \beta^{\mathrm{type}(v)}$. Consequently, slices belonging to frame $v$ and processed at speed $f_t^j \in \mathcal{F}$ have service time $\tau^v = w^v / f_t^j$, where $\tau^v$ is exponentially distributed with mean $\mathbf{E}\left[ \tau^v \mid f_t^j \right] = \beta^{\mathrm{type}(v)} / f_t^j$. Due to the memoryless property of the exponential distribution, if a slice belonging to frame $v$ is scheduled on processor $j$ at time $t$, then it will finish decoding in time slot $t$ (i.e. in $\Delta t$ seconds) with probability $\theta^v\left(f_t^j\right) = 1 - \exp\left(-\dfrac{f_t^j}{\beta^{\text{type}(v)}}\Delta t\right)$, regardless of the number of times it was previously scheduled. In other words, if a slice takes multiple time slots to decode, then the memoryless property implies that it is not necessary to know the number of cycles that were spent decoding the slice in past time slots to predict the distribution of remaining cycles. Hence, assuming exponentially distributed service times greatly reduces the number of states required in our Markovian traffic model (see Appendix A for more details). This is (implicitly) why a lot of prior research on power management using MDPs assumes exponential service times (e.g. [17] [18]).

We note that, if a slice finishes decoding before the time quantum is up, then we start decoding another slice (from the same frame) during the "slack" time, which is the time between the beginning of the next time quantum and the time that the originally scheduled slice finished decoding. We discuss this in more detail in Section 3.2.

## 2.5. State evolution and system dynamics

To fully characterize the video traffic, we need to understand how the traffic state $\mathcal{T}_t = \left(\mathcal{C}_t, \mathbf{x}_t, \mathbf{r}_t\right)$, comprising the current frame set $\mathcal{C}_t$, the buffer state $\mathbf{x}_t$, and the dependency state $\mathbf{r}_t$, evolves over time.

The transition of the current frame set from $\mathcal{C}_t$ to $\mathcal{C}_{t+1}$ is independent of the scheduling action; in fact, as illustrated in Fig. 2, it is deterministic and periodic for a fixed GOP structure, and therefore the sequence of current frame sets $\left\{\mathcal{C}_t \mid t \in \mathbb{N}\right\}$ can be modeled as a deterministic Markov chain.

Unlike the current frame set transition, the transition of the buffer state from $x_t^v$ to $x_{t+1}^v$ depends on the scheduling action and processor frequency. Let $z_t^{jv} = z_t^{jv}\left(f_t^j, y_t^{jv}\right)$ denote the number of slices belonging to frame $v$ that finish decoding on processor $j$ at time $t$. Note that $z_t^{jv} \leq y_t^{jv}$. For notational convenience, we define $\mathbf{Z}_t = \left[z_t^{jv}\right]_{jv}$, $\mathbf{z}_t^v = \left(z_t^{jv} \mid j \in \{1,\ldots,M\}\right)^T$, and $\mathbf{z}_t^j = \left(z_t^{jv} \mid v \in \mathcal{C}_t\right)$. Let $p_z\left(z_t^{jv} \mid f_t^j, y_t^{jv}\right)$ denote the probability that $z_t^{jv}$ slices are decoded on processor $j$ in time slot $t$ given the frequency $f_t^j$ and scheduling action $y_t^{jv}$: that is,

$$p_z\left(z_t^{jv} \mid f_t^j, y_t^{jv}\right) = \begin{cases} \theta^v\left(f_t^j\right), & y_t^{jv} = 1 \text{ and } z_t^{jv} = 1 \\ 1 - \theta^v\left(f_t^j\right), & y_t^{jv} = 1 \text{ and } z_t^{jv} = 0 \\ 1, & y_t^{jv} = 0 \text{ and } z_t^{jv} = 0 \\ 0, & \text{otherwise} \end{cases}$$

Before we can write the buffer recursion governing the transition from $x_t^v$ to $x_{t+1}^v$, we need to define a partition of the current frame set $\mathcal{C}_{t+1}$. The partition divides $\mathcal{C}_{t+1}$ into two sets: a set of frames that persist from time $t$ to $t+1$ because they have display deadlines $d^{v,\mathrm{disp}} > t$, i.e., $\mathcal{C}_t \cap \mathcal{C}_{t+1}$; and, a set of newly arrived frames with arrival times $t^v = t + 1$, i.e., $\mathcal{C}_{t+1} \setminus \mathcal{C}_t \triangleq \mathcal{C}_{t+1} - \mathcal{C}_t \cap \mathcal{C}_{t+1}$. Based on this partition, $x_{t+1}^v$ can be determined from $x_t^v$ and $\sum_{j=1}^M z_t^{jv}$ as follows

$$x_{t+1}^v = \begin{cases} x_t^v - \sum_{j=1}^M z_t^{jv}, & \text{if } v \in \mathcal{C}_t \cap \mathcal{C}_{t+1} \\ l^v, & \text{if } v \in \mathcal{C}_{t+1} \setminus \mathcal{C}_t. \end{cases} \tag{1}$$

The sequence of buffer states $\left\{x_t^v \mid t \in \mathbb{N}\right\}$ can be modeled as a controlled Markov chain. Note that the buffer state for frame $v$, i.e. $x_t^v$, is only defined for $t \in \left[t^v, d^{v,\mathrm{disp}}\right]$. We will refer to this range of times as the *lifetime* of frame $v$.

The transition of the dependency state from $r_t^v$ to $r_{t+1}^v$ for $v \in \mathcal{C}_t \cap \mathcal{C}_{t+1}$ can be determined as follows:

$$r_{t+1}^v = \begin{cases} 1, & \text{if } x_t^u - \sum_{j=1}^M z_t^{ju} = 0 \text{ for all } u \in \mathcal{C}_t \text{ such that } (u,v) \in \mathcal{E} \\ 1, & \text{if } r_t^v = 1 \\ 0, & \text{otherwise.} \end{cases} \tag{2}$$

The first line in (2) states that frame $v$ can be decoded in time slot $t+1$ if all of its parents are completely decoded at the end of time slot $t$. The second line in (2) states that if frame $v$ can be decoded in time slot $t$ then it can also be decoded in time slot $t+1$. Meanwhile, the initial value of dependency state $r_{t+1}^v$ for $v \in \mathcal{C}_{t+1} \setminus \mathcal{C}_t$ can be determined as follows:

$$r_{t+1}^v = \begin{cases} 1, & \text{if } x_t^u - z_t^u = 0 \text{ for all } u \in \mathcal{C}_t \text{ such that } (u,v) \in \mathcal{E} \\ 0, & \text{otherwise} \end{cases} \tag{3}$$

It follows from (2) that the sequence of dependency states $\left\{r_t^v \mid t \in \mathbb{N}\right\}$ can be modeled as a controlled Markov chain. Note that, similar to the buffer state, the dependency state is only defined for the lifetime $t \in \left[t^v, d^{v,\mathrm{disp}}\right]$. Note that (2) and (3) imply that, if frame $v$ is an I frame, then $r_{v,t} = 1$ for the frame's entire lifetime.

Because the individual components of the traffic state $\mathcal{T}_t^i \triangleq \left( \mathcal{C}_t^i, \mathbf{x}_t^i, \mathbf{r}_t^i \right)$ evolve as controlled Markov chains, the sequence of traffic states $\{ \mathcal{T}_t | \ t \in \mathbb{N} \}$ can be modeled as a controlled Markov chain.

### 2.6. Power cost and slice decoding rate

The power-frequency function $\rho \left( f_t^j \right)$ maps the $j$th processor's speed $f_t^j$ to its expected power consumption (watts). We assume that the power-frequency function is a strictly convex and increasing function of the frequency $f$ and that it is the same for each processor. We also consider the expected power consumed by the instruction, data, and L2 cache using a function $\sigma \left( f_t^j, y_t^{jv}, \mathrm{type} \left( v \right) \right)$, which maps the $j$th processor's speed $f_t^j$, the scheduling action $y_t^{jv}$, and frame type $\mathrm{type} \left( v \right)$ to power consumption (watts). Thus, the total expected power consumed by processor $j$ (and the associated accesses to the various caches) at time $t$ can be written as

$$P \left( \mathcal{C}_t, f_t^j, \mathbf{y}_t^j \right) = \rho \left( f_t^j \right) + \sum_{v \in \mathcal{C}_t} \sigma \left( f_t^j, y_t^{jv}, \mathrm{type} \left( v \right) \right) \text{ (watts).} \qquad (4)$$

Note that different frame types require different cache access patterns so the cache access power depends on the frame type (i.e. $\sigma \left( f_t^j, y_t^{jv}, \mathrm{type} \left( v \right) \right)$ depends on $\mathrm{type} \left( v \right)$). In Section 0, we describe how we populate (4) by profiling an H.264 video decoder on our MPARM simulation platform.

We consider the following QoS metric in each time slot $t$:

$$Q \left( f_t^j, y_t^{jv}, \mathrm{type} \left( v \right) \right) = \sum_{z_t^{jv} \leq y_t^{jv}} p_z \left( z_t^{jv} \mid f_t^j, y_t^{jv} \right) z_t^{jv} \qquad (5)$$

This QoS metric is simply the expected number of slices belonging to frame $v$ that will be decoded on processor $j$ in time slot $t$. We will refer to (5) as the *slice decoding rate* for frame $v$ on processor $j$. For notational simplicity, in the remainder of the report, we will omit the functional dependence of (4) and (5) on $\mathrm{type} \left( v \right)$.

### 2.7. Markov decision process formulation

In this subsection, we formulate the problem of energy-efficient slice-parallel video decoding on $M$ processors. In each time slot $t$, the objective is to determine the scheduling action $y_t^{jv}$, for all $j \in \left\{ 1, 2, \ldots, M \right\}$ and $v \in \mathcal{C}_t$, and the frequency vector $\mathbf{f}_t$, in order to minimize the total average power consumption subject to a constraint on the average slice decoding rate. The total *discounted*[5] average power consumption and slice decoding rate can be expressed as

$$\bar{P} = \mathbf{E}\left[\sum_{t=0}^{\infty}\sum_{j=1}^{M}\gamma^t P\left(\mathcal{C}_t, f_t^j, \mathbf{y}_t^j\right)\right], \text{ and} \tag{6}$$

$$\bar{Q} = \mathbf{E}\left[\sum_{t=0}^{\infty}\sum_{j=1}^{M}\sum_{v\in\mathcal{C}_t}\gamma^t Q\left(f_t^j, y_t^{jv}\right)\right], \tag{7}$$

respectively, where $\gamma\in\left[0,1\right)$ is the discount factor, and the expectation is over the sequences of traffic states $\{\mathcal{T}_t \mid t\in\mathbb{N}\}$. Stated more formally, the optimization objective and constraints are as follows:

$$\min_{\mathbf{f}_t,\mathbf{Y}_t,\;\forall t\in\mathbb{N}} \bar{P}$$

**Subject to** :

Slice decoding rate constraint: $\bar{Q} \geq \bar{\eta}$

Buffer constraint: $\sum_{j=1}^{M} y_t^{jv} \leq x_t^v, \;\forall v\in\mathcal{C}_t, \;\forall t\in\mathbb{N}$

Processor constraint: $\sum_{v\in\mathcal{C}_t} y_t^{jv} \leq 1, \;\forall j\in\left\{1,\ldots,M\right\}, \;\forall t\in\mathbb{N}$

Dependency constraint: if $r_t^v = 0$, then $\sum_{j=1}^{M} y_t^{jv} = 0, \;\forall v\in\mathcal{C}_t, \;\forall t\in\mathbb{N}$ 

$$\tag{8}$$

where $\bar{\eta}$ is the discounted slice decoding rate constraint. Note that it is a trivial extension to maximize the average slice decoding rate under an average power constraint.

The constrained optimization defined in (8) can be formulated as an unconstrained MDP by introducing a Lagrange multiplier $\lambda\in\mathbb{R}_+$ associated with the slice decoding rate constraint. Note that the buffer, processor, and dependency constraints defined in (8) must still hold in every time slot, however, for notational simplicity, we will omit them from our exposition in the remainder of the report. We can define the Lagrangian cost function:

$$c_\lambda\left(\mathcal{C}_t, \mathbf{f}_t, \mathbf{Y}_t\right) = \sum_{j=1}^{M} P\left(\mathcal{C}_t, f_t^j, \mathbf{y}_t^j\right) - \lambda\left(\sum_{j=1}^{M}\sum_{v\in\mathcal{C}_t}\gamma^t Q\left(f_t^j, y_t^{jv}\right) - \bar{\eta}\right), \tag{9}$$

For a fixed $\lambda$, in each time slot $t$, the unconstrained problem's objective is to determine the frequency vector $\mathbf{f}_t$ and scheduling matrix $\mathbf{Y}_t$ in order to minimize the average Lagrangian cost. The discounted average Lagrangian cost can be expressed as

$$L_\lambda = \min_{\mathbf{f}_t,\mathbf{Y}_t,\;\forall t\in\mathbb{N}} \mathbf{E}\left[\sum_{t=0}^{\infty}\gamma^t c_\lambda\left(\mathcal{C}_t, \mathbf{f}_t, \mathbf{Y}_t\right)\right] \tag{10}$$

---

[5] In this report, for mathematical convenience, we use *discounted* averages instead of conventional averages; however, the problem can be formulated using non-discounted averages. We refer the interested reader to [17] for an intuitive justification for using discounted averages.

Letting $p\left(\mathcal{T}' \mid \mathcal{T}, \mathbf{f}, \mathbf{Y}\right)$ denote the traffic state transition probability function, the problem of minimizing (10) can be mapped to the following dynamic programming equation:

$$V_\lambda^*\left(\mathcal{T}\right) = \min_{\mathbf{f}, \mathbf{Y}} \left\{ c_\lambda\left(\mathcal{C}, \mathbf{f}, \mathbf{Y}\right) + \gamma \sum_{\mathcal{T}'} p\left(\mathcal{T}' \mid \mathcal{T}, \mathbf{f}, \mathbf{Y}\right) V_\lambda^*\left(\mathcal{T}'\right) \right\}, \tag{11}$$

which can be solved using the well-known value iteration algorithm [14] as follows:

$$V_{n+1,\lambda}\left(\mathcal{T}\right) = \min_{\mathbf{f}, \mathbf{Y}} \left\{ c_\lambda\left(\mathcal{T}, \mathbf{f}, \mathbf{Y}\right) + \gamma \sum_{x'} p\left(\mathcal{T}' \mid \mathcal{T}, \mathbf{f}, \mathbf{Y}\right) V_{n,\lambda}\left(\mathcal{T}'\right) \right\} \tag{12}$$

where $n$ is the iteration index, $V_{0,\lambda}\left(\mathcal{T}\right)$ is initialized to 0 for all $\mathcal{T}$, and $V_{n,\lambda}\left(\mathcal{T}\right)$ approaches $V_\lambda^*\left(\mathcal{T}\right)$ as $n \rightarrow \infty$ [14].

## 3. LOW COMPLEXITY SOLUTION

Unfortunately, solving (12) directly is a computationally intractable problem for two reasons. First, the number of traffic states exponentially increases with the number of frames in the current frame set. Second, the action-space exponentially increases with the number of processors $M$ because $\mathbf{f} \in \mathcal{F}^M$ and, accounting for the processor constraint defined in Section 2.4 and the fact that all processors are homogeneous[6], we have to consider at most $2^M = \left|\left\{0, 1\right\}^M\right|$ scheduling actions $\mathbf{Y}$. The following examples demonstrate the explosion of the state and action spaces.

**Example 3: Exponential growth of the state space:** *Consider a decoding workload $D_1$ with the GOP structure in Fig. 2 and the four current frame sets listed in Example 1. Assume that there are $l_1 = 4$ slices per frame. For this workload, there are $2 \cdot 4^3 + 2 \cdot 4^4 = 640$ potential traffic states (because there are two current frame sets with 3 frames and two with 4 frames). Now, consider a decoding workload $D_2$ with the same GOP structure and current frame sets, but with $l_2 = 8$ slices per frame. For this workload, there are $2 \cdot 8^3 + 2 \cdot 8^4 = 9216$ potential traffic states.*

**Example 4: Exponential growth of the action space:** *Consider a system $S_1$ with $M_1 = 4$ processors and $\left|\mathcal{F}_1\right| = 4$ frequencies available for each processor. This system has $4^4 = 256$ possible frequency configurations and $2^4 = 16$ possible scheduling actions, for a total of $4^4 \times 2^4 = 2^{12}$ actions. Now, consider a*

---

[6] The homogeneity assumption means that all processors have the same cost function and the same set of available operating frequencies. It implies that the MDP only needs to determine whether or not a slice is scheduled, and then slices can be greedily assigned to processors. Since at most $M$ slices can be scheduled in each time slot (i.e. one slice per processor), this implies that $\mathbf{Y}$ reduces to $M$ binary decisions that must be made jointly.

*system $S_2$ with $M_2 = 8$ processors and $\left| \mathcal{F}_2 \right| = 4$ frequencies available for each processor. This system has*

*$4^8 = 2^{16}$ possible frequency configurations and $2^8$ possible scheduling actions, for a total of $4^8 \times 2^8 = 2^{24}$*

*actions.*

Clearly, the reason for the exponential growth in the state space (respectively, action space) is that the optimization simultaneously considers the states (respectively, scheduling actions and processor frequencies) of multiple frames. However, carefully studying the optimization objective and constraints defined in (8), it is clear that the only reason these need to be optimized jointly is the processor constraint, which ensures that only one slice is assigned to each processor in each time slot. Motivated by this weak coupling among tasks, we propose a two-level scheduler to approximately solve (8): The first-level scheduler determines the optimal scheduling actions and processor frequencies for each frame under the assumption that each frame has exclusive access to the $M$ processors. Given the results of the first-level scheduler, the second-level scheduler determines the final slice-to-processor and frequency-to-processor mappings.

### 3.1. First-level scheduler

The first-level scheduler computes a value function $V^v\left( \mathcal{C}, x^v, r^v \right)$ for every frame in a GOP. This value function only depends on the current frame set, the frame's buffer state $x^v$, and the frame's dependency state $r^v$. Note that the current frame set indicates the remaining lifetime of a frame and describes the connections to its parents and children. Hence, the current frame state will have a significant impact on the optimal scheduling and DVFS decisions for the frame. To account for the dependencies among frames, we define the $v$th frame's value function $V^v\left( \mathcal{C}, x^v, r^v \right)$ in such a way that it includes the values of its children. In this way, frames with many children (e.g. I frames) can account for how their scheduling and frequency decisions impact the future performance of their children. We describe the first-level scheduler in more detail in the remainder of this section.

#### 3.1.1. Frame-level value iteration

The first-level scheduler performs the frame-level value iteration algorithm illustrated in Table 3 to compute the optimal value functions $\left\{ V^{v,*} : v \in \mathcal{V}^g \right\}$. Similar to the conventional value iteration algorithm [14], the proposed frame-level value iteration algorithm iteratively updates the value functions for every state until a stopping condition is met. However, unlike the conventional value iteration algorithm, the proposed algorithm has multiple *coupled* value functions that need to be updated. Note that the coupling exists because the value of a frame depends on the values of its children. Due to this coupling, the form of the value function update (lines 5-9 in Table 3) is different from the conventional value iteration algorithm.

If it is not possible to make any decisions for a frame in the current traffic state, then we set the frame's value to 0 in that state. Hence, if a frame is not in the current frame set (i.e. $v \notin \mathcal{C}$), does not have its dependencies satisfied (i.e. $r^v = 0$), or is in the current frame set, but is already fully decoded (i.e. $v \in \mathcal{C}$ and $x^v = 0$), then we set the frame's value to 0 (line 8 in Table 3). The more interesting case is when the frame is in the current frame set, still has undecoded slices, and has its dependencies satisfied (i.e. $v \in \mathcal{C}$, $x^v > 0$, and $r^v = 1$). In this case, the value function update comprises four distinct terms: the power consumed by each processor in the current state; the expected slice decoding rate on each processor in the current state; the expected future value of frame $v$; and the sum of the expected future values of the $v$th frame's children. Note that the expected future value of frame $v$, i.e. $\gamma V_{n,\lambda}^v \left( \mathcal{C}', x^v - \left\| \mathbf{z}^{1:M,v} \right\|_1, r^{v\prime} \right)$, is 0 if $v \notin \mathcal{C}'$; and, the sum of the expected future values of the children's frames, i.e. $\gamma \sum_{u \in \mathcal{C}' : v \prec u, \ r^{u\prime} = 1} V_{n,\lambda}^u \left( \mathcal{C}', l^{u\prime}, r^{u\prime} \right)$, is 0 if $x^v - \left\| \mathbf{z}^{1:M,v} \right\|_1$ is not 0 (because $x^v - \left\| \mathbf{z}^{1:M,v} \right\|_1$ must be 0 for $r^{v\prime}$ to be 1). In other words, the parent frame's value function is coupled with the children's value functions only if the parent frame gets fully decoded.

The following example illustrates the reduction in state-space size achieved by applying the frame-level value iteration algorithm.

**Example 5: Reduction in state space size:** *Consider a decoding workload $D_1$ with the GOP structure in Fig. 2 and the four current frame sets listed in Example 1. Assume that there are $l_1 = 4$ slices per frame. For this workload, the I and P frame each require $4 \cdot 4 \cdot 2 = 32$ states because they are in all four current frames sets. Meanwhile, the B frames require $3 \cdot 4 \cdot 2 = 24$ states because they are in only three of the current frame sets. Now, consider a decoding workload $D_2$ with the same GOP structure and current frame sets, but with $l_2 = 8$ slices per frame. For this workload, there are $4 \cdot 8 \cdot 2 = 64$ states for the I and P frames and $3 \cdot 8 \cdot 2 = 48$ states for each B frame.*

**Table 3. Frame-level value iteration algorithm performed by the first-level scheduler.**

| | |
|---|---|
| **1.** | **Initialize:** $V_{0,\lambda}^v\left(\mathcal{C}, x^v, r^v\right) = 0$ for all $v \in \mathcal{V}^g$, $\mathcal{C}$, $x^v \in \left\{0, \ldots, l^v\right\}$, and $r^v \in \left\{0, 1\right\}$ |
| **2.** | **Repeat** |
| **3.** | $\Delta \leftarrow 0$ |
| **4.** | **For each** $v \in \mathcal{V}^g$, $\mathcal{C}$, $x^v \in \left\{0, \ldots, l^v\right\}$, and $r^v \in \left\{0, 1\right\}$ |
| **5.** | **If** $v \in \mathcal{C}$, $x^v > 0$, and $r^v = 1$ (i.e. frame $v$ is in the current frame set, still has undecoded slices, and has its dependencies satisfied) |
| **6.** | $$V_{n+1,\lambda}^v\left(\mathcal{C}, x^v, r^v\right) = \min_{\mathbf{f}^{1:M,v}, \mathbf{y}^{1:M,v}} \left\{ \begin{array}{l} \sum_{j=1}^M \left[\rho\left(f^{jv}\right) + \sigma\left(f^{jv}, y^{jv}\right) - \lambda Q\left(f^{jv}, y^{jv}\right)\right] \\ \gamma \sum_{\mathbf{z}^{1:M,v} \leq \mathbf{y}^{1:M,v}} \prod_{j=1}^M p_z\left(z^{jv} \mid f^{jv}, y^{jv}\right) \left[\begin{array}{l} V_{n,\lambda}^v\left(\mathcal{C}', x^v - \left\|\mathbf{z}^{1:M,v}\right\|_1, r^{v\prime}\right) + \\ \sum_{\substack{u \in \mathcal{C}': v \prec u \\ r^{u\prime} = 1}} V_{n,\lambda}^u\left(\mathcal{C}', l^u, r^{u\prime}\right) \end{array}\right] \end{array}\right\}$$ (13) |
| **7.** | **Else** |
| **8.** | $V_{n+1,\lambda}^v\left(\mathcal{C}, x^v, r^v\right) = 0$ |
| **9** | **End** |
| **10.** | **End** |
| **11.** | $\Delta \leftarrow \max\left(\Delta, \left\|V_{n+1,\lambda}^v\left(\mathcal{C}, x^v, r^v\right) - V_{n,\lambda}^v\left(\mathcal{C}, x^v, r^v\right)\right\|\right)$ |
| **12.** | $n \leftarrow n + 1$ |
| **13.** | **Until** $\Delta < \epsilon$ (a small positive number) |
| **14.** | **Output:** $\left\{V^{v,*} : v \in \mathcal{V}^g\right\}$ |

3.1.2. Decomposing the monolithic frame-level value iteration update

The frame-level value iterations allow us to eliminate the exponential growth of the state space with respect to the number of frames in the current frame set, but we still have to address the fact that the optimization in (13) (Line 6 of Table 3) requires a search over an exponential number of scheduling and frequency vectors. In this subsection, we discuss how to decompose the monolithic update defined in (13) into $M$ stages (hereafter, *sub-value iterations*), each corresponding to a local scheduling problem on a single processor. These $M$ sub-value iterations can be performed iteratively, using the output of the $j$th processor's sub-value iteration as the input to the $(j-1)$st processor's sub-value iteration. Importantly, decomposing the monolithic update into $M$ sub-value iterations significantly reduces the computational complexity of the update. The decomposition is illustrated in Fig. 4 and described in detail in the remainder of this subsection. In Appendix B, we discuss the complexity of the frame-level value iteration algorithm with decomposed value iteration update.

Let $\mathbf{f}^{1:j,v}$, $\mathbf{y}^{1:j,v}$, and $\mathbf{z}^{1:j,v}$, for $1 < j \leq M$, be vectors denoting the frequencies, scheduling actions, and number of decoded slices for frame $v$ on processors 1 through $j$. Let $\left\| \mathbf{z}^{1:j,v} \right\|_1 = \sum_{i=1}^{j} z^{iv}$ denote the $\ell_1$-norm of the vector $\mathbf{z}^{1:j,v}$. Let $\mathbf{E}_{\mathbf{z}^{1:j,v} \mid \mathbf{f}^{1:j,v}, \mathbf{y}^{1:j,v}} \left[ g\left( \mathbf{z}^{1:j,v} \right) \right] = \sum_{\mathbf{z}^{1:j,v} \leq \mathbf{y}^{1:j,v}} \prod_{i=1}^{j} p_z \left( z^{iv} \mid f^{iv}, y^{iv} \right) \left[ g\left( \mathbf{z}^{1:j,v} \right) \right]$ be shorthand for taking the expectation of a function $g\left( \mathbf{z}^{1:j,v} \right)$ with respect to the distribution of decoded slices on processors 1 through $j$. Equipped with this new notation, we derive the sub-value iterations from (13).

Notice that, in (13), $\rho\left( f^{Mv} \right) + \sigma\left( f^{Mv}, y^{Mv} \right) - \lambda Q\left( f^{Mv}, y^{Mv} \right)$ is independent of $\mathbf{z}^{1:M-1,v}$, and the expression $\sum_{\mathbf{z}^{1:M-1,v} \leq \mathbf{y}^{1:M-1,v}} \prod_{j=1}^{M-1} p_z \left( z^{jv} \mid f^{jv}, y^{jv} \right)$ sums to 1. Hence, we can rewrite (13) as follows:

$$
V_{n+1,\lambda}^{v}\left( \mathcal{C}, x^v, r^v \right) =
$$

$$
\min_{\substack{\mathbf{f}^{1:M-1,v} \\ \mathbf{y}^{1:M-1,v}}} \left\{ 
\begin{array}{l}
\sum_{i=1}^{M-1} \left[ \rho\left( f^{iv} \right) + \sigma\left( f^{iv}, y^{iv} \right) - \lambda Q\left( f^{iv}, y^{iv} \right) \right] + \\
\mathbf{E}_{\mathbf{z}^{1:M-1,v} \mid \mathbf{f}^{1:M-1,v}, \mathbf{y}^{1:M-1,v}} \underbrace{ \min_{\substack{f^{Mv} \\ y^{Mv}}} \left[ 
\begin{array}{l}
\rho\left( f^{Mv} \right) + \sigma\left( f^{Mv}, y^{Mv} \right) - \lambda Q\left( f^{Mv}, y^{Mv} \right) + \\
\gamma \mathbf{E}_{z^{Mv} \mid f^{Mv}, y^{Mv}} \left[ V_{n,\lambda}^{v}\left( \mathcal{C}', x^v - \left\| \mathbf{z}^{1:M,v} \right\|_1, r^{v\prime} \right) + \sum_{\substack{u \in \mathcal{C}': v \prec u \\ r^{v\prime}=1}} V_{n,\lambda}^{u}\left( \mathcal{C}', l^{u\prime}, r^{u\prime} \right) \right]
\end{array}
\right]}_{V_{n,\lambda}^{M-1,v}\left( \mathcal{C}, x^v, r^v \mid \left\| \mathbf{z}^{1:M-1,v} \right\|_1 \right)}
\end{array}
\right\} \quad (14)
$$

The inner minimization in (14) is the $M$ th processor's sub-value iteration, the result of which we denote by $V_{n,\lambda}^{M-1,v}\left( \mathcal{C}, x^v, r^v \mid \left\| \mathbf{z}^{1:M-1,v} \right\|_1 \right)$.

---

**Sub-value iteration at processor $M$ :**

$$
V_{n,\lambda}^{M-1,v}\left( \mathcal{C}, x^v, r^v \mid \left\| \mathbf{z}^{1:M-1,v} \right\|_1 \right) =
$$
$$\quad (15)$$
$$
\min_{f^{Mv}, \, y^{Mv}} \left\{ 
\begin{array}{l}
\rho\left( f^{Mv} \right) + \sigma\left( f^{Mv}, y^{Mv} \right) - \lambda Q\left( f^{Mv}, y^{Mv} \right) + \\
\gamma \mathbf{E}_{z^{Mv} \mid f^{Mv}, y^{Mv}} \left[ V_{n,\lambda}^{v}\left( \mathcal{C}', x^v - \left\| \mathbf{z}^{1:M-1,v} \right\|_1 - z^{Mv}, r^{v\prime} \right) + \sum_{u \in \mathcal{C}': v \prec u, \, r^{v\prime}=1} V_{n,\lambda}^{u}\left( \mathcal{C}', l^{u\prime}, r^{u\prime} \right) \right]
\end{array}
\right\}
$$

---

The $M$ th processor's sub-value iteration estimates the value of being in traffic state $\mathcal{T}^v = \left( \mathcal{C}, x^v, r^v \right)$ conditioned on processors 1 through $M-1$ successfully decoding $\left\| \mathbf{z}^{1:M-1} \right\|_1 \in \left\{ 0, 1, \ldots, M-1 \right\}$ slices. This value is calculated as the sum of (i) the immediate cost incurred by processor $M$ for processing slices belonging to frame $v$, i.e. $\rho\left( f^{Mv} \right) + \sigma\left( f^{Mv}, y^{Mv} \right) - \lambda Q\left( f^{Mv}, y^{Mv} \right)$, (ii) the expected discounted future value of frame $v$ transitioning to state $\mathcal{T}^{v\prime} = \left( \mathcal{C}', x^v - \left\| \mathbf{z}^{1:M-1,v} \right\|_1 - z^{Mv}, r^{v\prime} \right)$, i.e.

$\gamma \mathbf{E}_{z^{Mv}|f^{Mv},y^{Mv}}\left[V_{n,\lambda}^{v}\left(\mathcal{C}',x^{v}-\left\|\mathbf{z}^{1:M-1,v}\right\|_{1}-z^{Mv},r^{v}{}'\right)\right]$, and (iii) the expected discounted future value of the $v$ th frame's children, i.e. $\gamma \mathbf{E}_{z^{Mv}|f^{Mv},y^{Mv}}\left[\sum_{u\in\mathcal{C}':v\prec u,\ r^{u}{}'=1}V_{n,\lambda}^{u}\left(\mathcal{C}',l^{u}{}',r^{u}{}'\right)\right]$. The output of the $M$ th processor's sub-value iteration, i.e.

$$\left\{V_{n,\lambda}^{M-1,v}\left(\mathcal{C},x^{v},r^{v}\mid\left\|\mathbf{z}^{1:M-1,v}\right\|_{1}\right):x^{v}\in\left\{0,\dots,l^{v}\right\},\ \left\|\mathbf{z}^{1:M-1,v}\right\|_{1}\in\left\{0,1,\dots,M-1\right\}\right\},$$

is used as input to the $\left(M-1\right)$ st processor's sub-value iteration derived below. These outputs are represented by the rightmost nodes in Fig. 4.

To derive the sub-value iterations at processors $j\in\left\{2,\dots,M-1\right\}$, we first observe that $\rho\left(f^{M-1,v}\right)+\sigma\left(f^{M-1,v},y^{M-1,v}\right)-\lambda Q\left(f^{M-1,v},y^{M-1,v}\right)$ is independent of $\mathbf{z}^{1:M-2,v}$ and that the expression $\sum_{\mathbf{z}^{1:M-2,v}\leq\mathbf{y}^{1:M-2,v}}\prod_{j=1}^{M-2}p_{z}\left(z^{jv}\mid f^{jv},y^{jv}\right)$ sums to 1. Hence, similar to how we obtained (14) from (13), we can rewrite (14) as follows:

$$V_{n+1,\lambda}^{v}\left(\mathcal{C},x^{v},r^{v}\right)=$$
$$\min_{\substack{\mathbf{f}^{1:M-2,v}\\\mathbf{y}^{1:M-2,v}}}\left\{\sum_{i=1}^{M-2}\left[\rho\left(f^{iv}\right)+\sigma\left(f^{iv},y^{iv}\right)-\lambda Q\left(f^{iv},y^{iv}\right)\right]+\mathbf{E}_{\mathbf{z}^{1:M-2,v}|\mathbf{f}^{1:M-2,v},\mathbf{y}^{1:M-2,v}}\underbrace{\left[\min_{\substack{f^{M-1,v}\\y^{M-1,v}}}\left\{\begin{matrix}\rho\left(f^{M-1,v}\right)+\sigma\left(f^{M-1,v},y^{M-1,v}\right)-\lambda Q\left(f^{M-1,v},y^{M-1,v}\right)+\\\mathbf{E}_{z^{M-1,v}|f^{M-1,v},y^{M-1,v}}\left[V_{n,\lambda}^{M-1,v}\left(\mathcal{C},x^{v}-\left\|\mathbf{z}^{1:M-1,v}\right\|_{1},r^{v}\mid\left\|\mathbf{z}^{1:M-1}\right\|_{1}\right)\right]\end{matrix}\right\}\right]}_{V_{n,\lambda}^{M-2,v}\left(\mathcal{C},x^{v},r^{v}\mid\left\|\mathbf{z}^{1:M-2,v}\right\|_{1}\right)}\right\}\quad(16)$$

The inner minimization in (16) is the $\left(M-1\right)$ st processor's sub-problem, the result of which we denote by $V_{n,\lambda}^{M-2,v}\left(\mathcal{C},x^{v},r^{v}\mid\left\|\mathbf{z}^{1:M-2,v}\right\|_{1}\right)$. Repeating this process, we obtain the sub-value iterations for processors $2,\dots,M-1$:

---

**Sub-value iteration at processors** $j\in\left\{2,\dots,M-1\right\}$:

$$V_{n,\lambda}^{j-1,v}\left(\mathcal{C},x^{v},r^{v}\mid\left\|\mathbf{z}^{1:j-1,v}\right\|_{1}\right)=\min_{f^{jv},\ y^{jv}}\left\{\begin{matrix}\rho\left(f^{jv}\right)+\sigma\left(f^{jv},y^{jv}\right)-\lambda Q\left(f^{jv},y^{jv}\right)+\\\mathbf{E}_{z^{jv}|f^{jv},y^{jv}}\left[V_{n,\lambda}^{j,v}\left(\mathcal{C},x^{v}-\left\|\mathbf{z}^{1:j-1,v}\right\|_{1}-z^{jv},r^{v}\mid\left\|\mathbf{z}^{1:j,v}\right\|_{1}\right)\right]\end{matrix}\right\}.\quad(17)$$

---

The $j$ th processor's sub-value iteration estimates the value of being in traffic state $\mathcal{T}^{v}=\left(\mathcal{C},x^{v},r^{v}\right)$ conditioned on processors 1 through $j-1$ successfully decoding $\left\|\mathbf{z}^{1:j-1,v}\right\|_{1}\in\left\{0,1,\dots,j-1\right\}$ slices. This value is calculated as the sum of the immediate cost incurred by processor $j$, i.e. $\rho\left(f^{jv}\right)+\sigma\left(f^{jv},y^{jv}\right)-\lambda Q\left(f^{jv},y^{jv}\right)$, and an expectation over the value calculated by the $\left(j+1\right)$ st processor's sub-value iteration, i.e. $\mathbf{E}_{z^{jv}|f^{jv},y^{jv}}\left[V_{n,\lambda}^{j,v}\left(\mathcal{C},x^{v}-\left\|\mathbf{z}^{1:j-1,v}\right\|_{1}-z^{jv},r^{v}\mid\left\|\mathbf{z}^{1:j,v}\right\|_{1}\right)\right]$. The output of the $j$ th processor's sub-value iteration, i.e.

$$\left\{V_{n,\lambda}^{j-1,v}\left(\mathcal{C},x^{v},r^{v}\mid\left\|\mathbf{z}^{1:j-1,v}\right\|_{1}\right):x^{v}\in\left\{0,\dots,l^{v}\right\},\ \left\|\mathbf{z}^{1:j-1,v}\right\|_{1}\in\left\{0,1,\dots,j-1\right\}\right\},$$

is used as input to the $(j-1)$ st processor's sub-value iteration. These outputs are represented by the nodes in columns $j \in \{2, \ldots, M-1\}$ in Fig. 4.

Finally, using the same arguments as above, the sub-value iteration at processor $j = 1$ is defined as follows:

**Sub-value iteration at processor 1:**

$$V_{n+1,\lambda}^v\left(\mathcal{C}, x^v, r^v\right) = \min_{f^{1v},\, y^{1v}} \left\{ \rho\left(f^{1v}\right) + \sigma\left(f^{1v}, y^{1v}\right) - \lambda Q\left(f^{1v}, y^{1v}\right) + \mathbf{E}_{z^{1v}|f^{1v}, y^{1v}}\left[ V_{n,\lambda}^{1,v}\left(\mathcal{C}, x^v - z^{1v}, r^v \mid z^{1v}\right)\right] \right\}. \quad (18)$$

The output of the first processor's sub-value iteration includes (i) the immediate power costs incurred by all processors, (ii) the slice decoding rate of all processors, (iii) the expected discounted future value of frame $v$, and (iv) the expected future discounted value of frame $v$'s children. The output of the first processor's sub-value iteration during iteration $n$, i.e.

$$\left\{ V_{n+1,\lambda}^v\left(\mathcal{C}, x^v, r^v\right) : x^v \in \left\{0, \ldots, l^v\right\} \right\},$$

is used as input to the $M$ th processor's sub-value iteration during iteration $n+1$. These outputs are represented by the node in column 1 of Fig. 4.

Performing the $M$ sub-value iterations for frame $v$ on a single traffic state $\mathcal{T}^v = \left(\mathcal{C}, x^v, r^v\right)$ only requires a search over the (scalar) scheduling actions $y^{jv} \in \{0,1\}$ and frequencies $f^{jv} \in \mathcal{F}$ at each of the $O\left(M^2\right)$ nodes in Fig. 4. Therefore, using the proposed decomposition of the monolithic value function update significantly reduces the action-selection complexity.

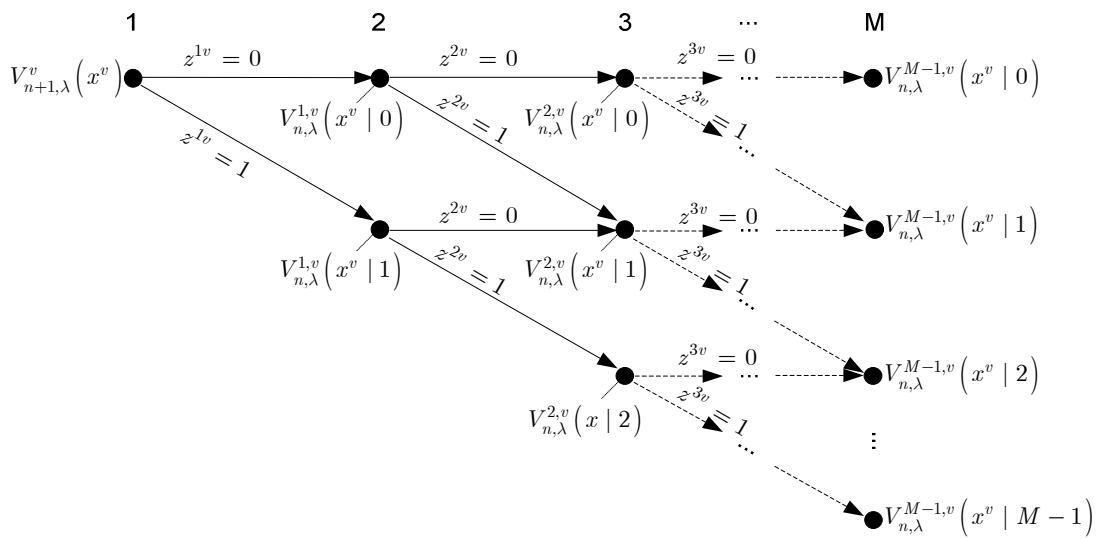

**Fig. 4. Decomposition of the monolithic value iteration update. For clarity, we omit the functional dependence of the value functions on the current frame set $\mathcal{C}$ and the dependency state $r^v$.**

### 3.1.3. Determining the approximately optimal policy

We define the *policy* $\pi^v : \left( \mathcal{C}, x^v, r^v \right) \rightarrow \left( \mathbf{f}^v, \mathbf{y}^v \right)$ as a mapping from the $v$th frame's traffic state $\mathcal{T}^v = \left( \mathcal{C}, x^v, r^v \right)$ to a pair of scheduling and frequency vectors $\left( \mathbf{f}^v, \mathbf{y}^v \right)$. If we know the optimal frame-level value functions $\left\{ V^{v,*} : v \in \mathcal{V}^g \right\}$, then we can determine the optimal action to take in each traffic state, and therefore the optimal policy, by finding the scheduling and frequency vectors that optimize (13). However, as we discussed earlier, this requires searching over an exponential number of scheduling and frequency vectors. Fortunately, it turns out that we can use the sub-value iterations proposed in Section 3.1.2 to find an approximately optimal policy. An algorithm for doing this is summarized in Table 4. In Appendix B, we discuss the complexity of determining the approximately optimal policy.

The key idea behind the algorithm in Table 4 is to find the (scalar) scheduling and frequency actions that optimize the sub-value functions defined in (15), (17), and (18) for each processor. However, there is one complication that must be dealt with before we can do this. Specifically, notice that the sub-value iterations for processors $j \in \left\{ 2, ..., M \right\}$ require knowledge of the number of slices that finish decoding on processors 1 through $j - 1$, i.e. $\left\| \mathbf{z}^{1:j-1,v} \right\|_1$. Unfortunately, we need to select the (scalar) scheduling action and processor frequency on processor $j$ before $\left\| \mathbf{z}^{1:j-1,v} \right\|_1$ is known. To work around this problem, the algorithm in Table 4 first selects the optimal (scalar) scheduling action and frequency for processor 1. Then, to select the optimal (scalar) scheduling actions and frequencies for processors $j \in \left\{ 2, ..., M \right\}$, the algorithm approximates $\left\| \mathbf{z}^{1:j-1,v} \right\|_1$ with the floor of its expected value, which depends on the optimal actions selected by processors 1 through $j - 1$, i.e. $\overline{Z}^{1:j-1,v,*} = \left\lfloor \mathbf{E}_{\mathbf{z}^{1:j-1,v} | \mathbf{f}^{1:j-1,v,*}, \mathbf{y}^{1:j-1,v,*}} \left[ \left\| \mathbf{z}^{1:j-1,v} \right\|_1 \right] \right\rfloor$.[7]

---

[7] The floor of $X$, denoted by $\left\lfloor X \right\rfloor$, is the largest integer value that is less than $X$.

**Table 4. Determining an approximately optimal policy for frame $v$.**

| 1. | **Input:** $V_\lambda^{v,*}\left(\mathcal{C}, x^v, r^v\right)$ for all $v \in \mathcal{V}^g$ |
|---|---|
| 2. | **For each** $\mathcal{C}$, $x^v \in \left\{0,\dots,l^v\right\}$, and $r^v \in \left\{0,1\right\}$ |
| 3. | Obtain $\left(f^{1v,*}, y^{1v,*}\right)$ as the argument that maximizes the 1$^{\text{st}}$ processor's sub-value function (Eq. (18)). |
| 2. | **For each** $j \in \left\{2,\dots,M\right\}$ |
| 4. | Approximate $\left\|\mathbf{z}^{1:j-1,v}\right\|_1$ with $\overline{Z}^{1:j-1,v,*} = \left\lfloor \mathbf{E}_{\mathbf{z}^{1:j-1,v}\mid \mathbf{f}^{1:j-1,v,*}, \mathbf{y}^{1:j-1,v,*}}\left[\left\|\mathbf{z}^{1:j-1,v}\right\|_1\right]\right\rfloor$ |
| | Obtain $\left(f^{jv,*}, y^{jv,*}\right)$ as the argument that maximizes the $j$th processor's sub-value function (Eq. (15) or (17)) given the optimal future value. |
| 5. | **End** |
| 7. | $\pi^{v,*}\left(\mathcal{C}, x^v, r^v\right) \leftarrow \left(\mathbf{f}^{1:M,v,*}, \mathbf{y}^{1:M,v,*}\right)$ |
| 8. | **End** |
| 14. | **Output:** $\pi^{v,*}\left(\mathcal{C}, x^v, r^v\right)$ |

### 3.2. Second-level scheduler

Given the optimal policies calculated by the first-level scheduler (i.e. $\pi^{v,*}\left(\mathcal{C}, x^v, r^v\right)$, for all $v \in \mathcal{V}^g$), it is very likely that slices belonging to different frames in the current frame set will want to be scheduled on the same processor in the same time slot, thereby violating the processor constraint defined in (8). To avoid this problem, the second-level scheduler determines the final slice-to-processor and frequency-to-processor mappings using an Earliest Deadline First (EDF) policy. Specifically, frame $v^{j,*}$ gets scheduled on processor $j$ at frequency $f^{jv,*}$ if $v^{j,*}$ is the solution to the following optimization:

$$v^{j,*} = \arg\min_{v:y^{jv,*}\neq 0} d^{v,\text{dec}},\qquad(19)$$

where $d^{v,\text{dec}}$ is the frame's decoding deadline and ties are broken randomly.

In addition to ensuring that the processor constraint defined in (8) is satisfied, a key role of the second-level scheduler is to guarantee that, once scheduled on a processor, a slice remains on that processor until it is either completely decoded or it expires. Keeping a slice on one core prevents the system from having to migrate a slice decoding task from one processor to another, which can be expensive in terms of delay, memory bandwidth, and system energy.

Finally, if a slice finishes decoding before the first-level scheduler's time quantum is up, then the second-level scheduler will start decoding another slice (from the same frame) during the "slack" time, which is the time between the beginning of the next time quantum and the time that the originally scheduled slice finished decoding. This is analogous to how slack reclamation is commonly used in the power management literature (see, e.g., [8][10][11]). That is, typically, an amount of time is allocated to a task based on its worst-case execution time (analogous to the first-level scheduler's time quantum), and, if the task completes before that time (analogous to it finishing before the first-level scheduler's time quantum is up), then the remaining slack time is used to schedule the next task. If there are no schedulable slices available to use the slack, then, similar to [3], the second-level scheduler idles the processor at the lowest operating frequency so that we do not waste energy.

In Appendix B, we discuss the complexity of the second-level scheduler.

### 3.3. *Impact of modeling assumptions on the optimal policy*

In this section, we discuss two major assumptions that slightly increase the power consumption of our proposed algorithm relative to the optimal algorithm.

First, at the first-level scheduler (driven by the MDP model), we assume that only one slice can be decoded on each core in each time slot, despite the fact that the second-level scheduler allows additional slices to be processed during the slack time. This may cause the first-level scheduler to be slightly more aggressive in its selection of processor frequencies than it would be with a more accurate (and more complex) MDP model that accounts for multiple slices being scheduled on each processor in each time slot. Consequently, more power will be consumed on average than required by the workload.

Second, we assume that the slices have exponential complexity distributions. This has an interesting impact on the temporal selection of operating frequencies when decoding a slice. Suppose that we have $T$ time slots of duration $\Delta t$ seconds to decode one video slice with random complexity $W \geq 0$ (cycles). Under the exponential complexity model, the probability of decoding a task in any given time slot is a constant conditioned on the operating frequency (i.e. it is independent of how many cycles have been processed in previous time slots). During the first (respectively, last) time slots spent decoding a slice, the exponential model tends to overestimate (respectively, underestimate) the probability of decoding a slice relative to the true probability, and therefore tends to select lower (respectively, higher) operating frequencies than are optimal. Overall, these policies approximately average out in terms of cycles allocated to decoding the slice (relative to the true optimal policy), but end up using more power than necessary (due to the convexity of the power-frequency function). We discuss this in more detail in Appendix A.

### 4. EXPERIMENTS

In this section, we describe our experimental framework in detail and evaluate our proposed algorithm. We note that we did not have access to a decoder that supports the sophisticated level of slice parallel decoding that our algorithm is designed to exploit. Specifically, the available decoder implementation can decode slices belonging to the same frame in parallel, but it cannot decode slices from different frames in parallel. Because the latter capability is essential to our proposed algorithm, we use Matlab to evaluate it, instead of the MPARM simulator.

*4.1. Experimental framework*

In order to validate our optimized multi-core scheduling approach in Matlab, we use accurate profiling/statistics generated from a parallelized H.264 decoder executed on a very sophisticated multiprocessor virtual platform simulator. In fact, in this work, we have extended and customized the multiprocessor ARM (MPARM) virtual platform simulator [15], which is a complete SystemC simulation environment for MPSoC architectural design and exploration. MPARM provides cycle-accurate and bus signal-accurate simulation for different processors. In our experiments, we have used the ARM9 Instruction Set Simulator as the main core. In addition, we have customized into MPARM its DVFS figures, number of cores, memory latencies, and cache size in order to accurately calculate and report the energy and power consumption of the cores and the different memory cache levels for our multimedia benchmark (i.e., H.264).

In order to run the H.264 decoder for up to CIF resolution on an ARM9 core, we have generated a specific experimental setup. In this experimental platform, we have integrated five ARM 9 cores running at a maximum frequency of 500MHz with DVFS support for each core (125MHz, 166MHz, 250MHZ at 1.07V and 500MHZ at 1.6V). These multiple processing cores replace the co-processing units, namely, the GPU, the DSP and the hardware acceleration featured in recent MPSoC models. Each of the processing cores has private 32KB L1 instruction and data caches. Moreover, we have also integrated 512KB of L2 cache memory that is shared between all the cores and connected to the main memory via an AMBA interconnection bus. The main memory is divided into private memory and shared memory. The private memory is L1 and L2 cacheable and the shared memory is only L2 cacheable. The synchronization between different cores is implemented with semaphores. The hardware configuration of our MPARM-based virtual platform is illustrated in Fig. 1 (Section 2).

We have used a real time operating system RTEMS (Real-Time Executive for Multiprocessor Systems) [16] in order to have multitasking execution on our MPARM multi-core experimental platform. Our optimized multi-core scheduling framework requires accurate statistics data output for each task (i.e., slice). Therefore, we have added an API that is able to create different interrupts from the application layer to the hardware layer for requesting a statistics record. This new API records, on select parts of the code, the execution time and the power consumption of the CPU, the instruction cache, the data cache, and the L2 cache. All of the statistics related to each task are then stored in a file.

For our multimedia benchmark, we have used the Joint Model reference software version 17.2 (JM 17.2) of an H.264 encoder. To support simple slice-level parallelism, we modified the H.264 decoder by allocating parts of the data to the shared memory instead of the private memory such that it is accessible from all the cores. We have then implemented our own memory management API (i.e., malloc, calloc, and free) for the MPARM shared memory. Finally, we have added a few instructions in the decoder code that tell us which part of the application is running on each core. We have then divided the decoder into three main tasks: the first task handles the parsing of the input video bit-stream to slices. This task is assigned to the master core (core 1). Then, the second task decodes the slices mapped by the master core. Slave cores process these slices in parallel. Finally, as a third task, the de-blocking filter is applied on the decoded slices. This last task was assigned to the master core. We use the developed API to record statistics for each task. Moreover, it also provides detailed statistics for each decoded slice, namely, the execution time, estimated power consumption figures, the slice index, the frame index, the GOP index and the assigned core. All these generated profiling data for each slice is then ported to Matlab and used as input into our algorithm to populate the expected power function and slice decoding complexity distributions. Since the MPARM experimental platform was only used to generate accurate profiling data, we have implemented a simple static scheduling algorithm to map the slices to the slave cores.

To generate our experimental results, we implemented the two-level scheduling algorithm proposed in Section 3 in Matlab. This algorithm, together with the slice-level data traces recorded from MPARM, allowed us to determine on-line scheduling and DVFS policies for the Silent and Foreman sequences (CIF resolution, 30 frames per second, 8 slices per frame) with an IBPB GOP structure as illustrated in Fig. 2. In our Matlab simulations, we assume a time slot duration of 1/90 s, which is one-third of the frame period. We divide each GOP into 12 current frame sets to capture the dependencies among frames. These 12 current frame sets are generated from the four unique current frame sets given in example 1 (Section 2.2) by repeating each for three consecutive time slots.[8] The system, application, and other parameters used in our experiments are given in Table 5. Importantly, although our MDP model assumes that the slice decoding complexities are exponentially distributed, we use the actual slice decoding times from the MPARM simulator when we simulate the scheduling and DVFS policies.

---

[8] In example 1 (Section 2.2), the time slot duration was equal to the frame duration (i.e. 1/30 s). Because we are now using a time slot duration equal to one-third of the frame duration (i.e. 1/90 s), we must repeat each of the current frame sets in example 1 three times.

**Table 5. Simulation parameters.**

| Parameter | Value(s) |
|---|---|
| No. slave cores ( $M$ ) | 1, 2, 4, 8 |
| Frequency set ( $\mathcal{F}$ ) | {125, 166, 250, 500} MHz |
| Sequence | Foreman (220 frames), Silent (300 frames) |
| Resolution | CIF (352 x 288) |
| GOP Structure | 'IBPB' |
| Frame rate | 30 frames per second |
| Time slot duration | 1/90 s |
| No. current frame sets | 12 |
| No. slices per frame | 8 |
| Lagrange multiplier ( $\lambda$ ) | 0, 50, 100, 200, 400, 800 |

*4.2. Trade-off between power consumption and Quality of Service*

The optimization proposed in (8) allows the system to trade-off power consumption and a QoS metric, namely, the slice decoding rate, which is roughly proportional to the frame rate. This trade-off can be made by adapting the Lagrange multiplier $\lambda$ in the cost function defined in (9). Intuitively, small values of $\lambda$ lead to scheduling and DVFS policies that favor power conservation over QoS, whereas larger values of $\lambda$ lead to policies that favor QoS over power consumption. Fig. 5 shows the trade-off between the average power consumption and average frame rate for the values of $\lambda$ given in Table 5 and $M = 1, 2, 4$, and 8 processors.

Fig. 5(a) and Fig. 5(b) show the average power consumption *per core* versus the average decoded frame rate for the Foreman and Silent sequences, respectively. The power-QoS pairs in the lower left of these two figures occur when $\lambda = 0$ and correspond to a scheduling policy that never schedules any tasks and a DVFS policy that always selects the lowest operating frequency, thereby achieving a QoS of zero frames per second. The minimum power consumption per core, which is approximately 20 mW, is due to leakage power. If we were to introduce DPM into our optimization framework, then this minimum power would be significantly lower. Clearly, as $\lambda$ increases, the QoS is improved at the expense of power; as the number of processors increases, less power is required per processor to decode at a given QoS; and, depending on the video source characteristics (e.g. Foreman vs. Silent), the achievable QoS varies for a given power consumption (in this case, Silent receives a higher QoS than Foreman for the same power consumption because Silent is a lower activity sequence).

Fig. 5(c) and Fig. 5(d) show the average *total* power consumption versus the average decoded frame rate for the Foreman and Silent sequences, respectively. It is interesting to note that, as the decoded frame rate decreases, having less processors results in less overall power consumption. This is due to the large leakage power incurred by each processor, which, as mentioned before, could be significantly reduced using DPM in addition to DVFS.

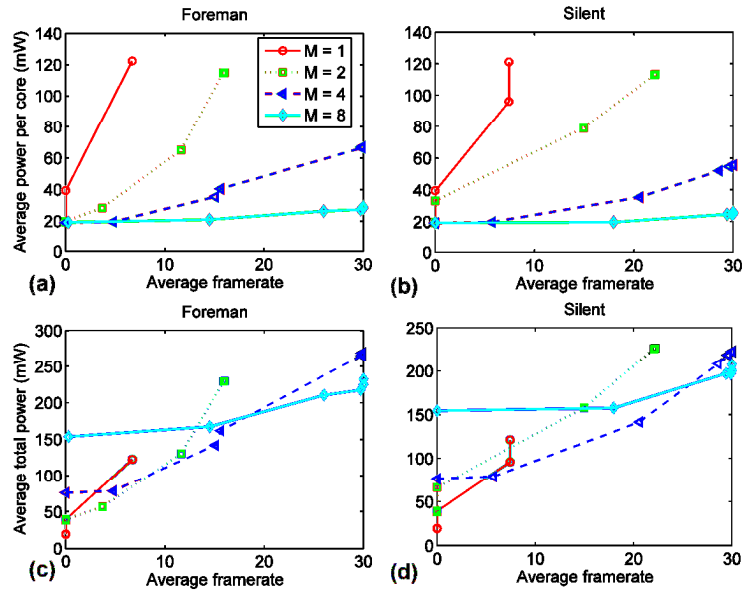

**Fig. 5. Power consumption versus the average decoded frame rate for CIF resolution sequences. (a,b) Average power consumption per core versus the average decoded frame rate. (c,d) Average total power consumption versus the average decoded frame rate.**

It is clear from Fig. 5 that the proposed scheduling algorithm exploits the loss-tolerant nature of video decoding tasks to achieve lower decoded frame rates when the energy-budget does not allow for full frame rate decoding. An important question is whether or not the algorithm could do significantly better. In the next subsection, to answer this question, we look at some statistics on which frames miss their deadlines most frequently.

*4.3. Display deadline miss rates*

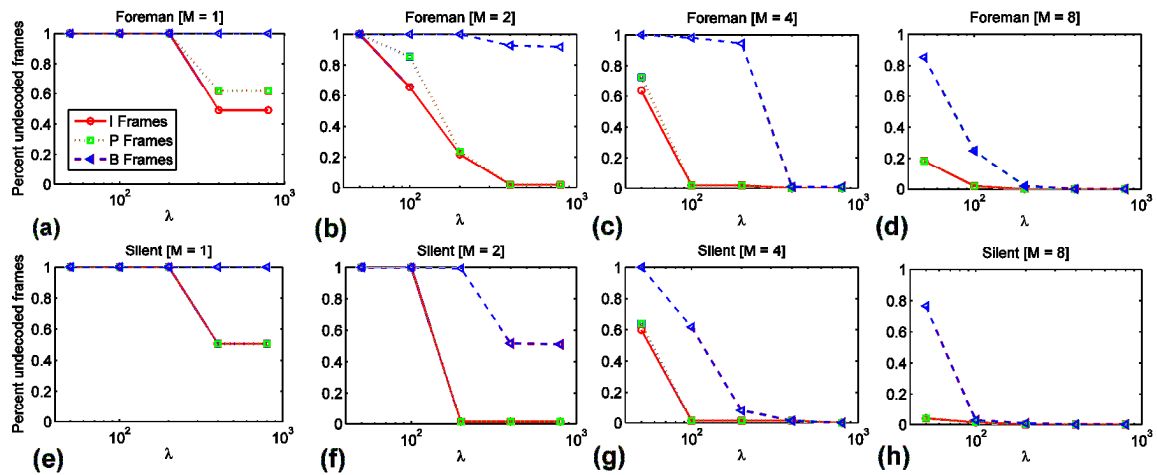

**Fig. 6. Fractions of I, P, and B frames that miss their display deadline as a function of the parameter $\lambda$ for CIF resolution sequencies. (a,b,c,d) Results for the Foreman sequence with M = 1, 2, 4, and 8 processors, respectively. (e,f,g,h) Results for the Silent sequence with M = 1, 2, 4, and 8 processors, respectively.**

Fig. 6 shows the fractions of I, P, and B frames that miss their display deadline as a function of the parameter $\lambda$ (for the values of $\lambda$ listed in Table 5). The results show that the proposed on-line scheduling and DVFS optimization has a very desirable property: as minimizing power becomes more important (i.e. $\lambda$ decreases), B frames are the first to miss their deadlines, followed by P frames, and then I frames. In other words, due to the smart scheduling algorithm, the QoS (i.e. frame rate) decreases slowly with the power consumption. In contrast, a scheduling policy that allows P frames to be lost before B frames, or I frames before P frames, is inherently suboptimal because a deadline miss by one I or P frame induces deadline misses of dependent frames, adversely impacting the QoS.

### 4.4. Impact of video resolution

The frame size will impact the system performance and power consumption in a several important ways. To demonstrate, we have included simulation results for QCIF resolution videos in Fig. 7 and Fig. 8, which complement the simulation results for CIF resolution sequences in Fig. 5 and Fig. 6. Fig. 7 shows the trade-off between the average power consumption and average frame rate for $\lambda \in \{0, 3, 6, 12, 25, 50, 100, 200, 400, 800, 1600, 3200, 6400\}$. Fig. 8 shows the fractions of I, P, and B frames that miss their display deadline as a function of $\lambda$. Comparing Fig. 7 and Fig. 8 (QCIF resolution) to Fig. 5 and Fig. 6 (CIF resolution), we observe:

1. **For a fixed number of cores and a fixed frame rate, increasing the video resolution requires higher power consumption.** Alternatively, for a fixed number of cores and a fixed power consumption level, increasing the video resolution will decrease the frame rate. This is because higher resolution frames have higher decoding complexity (proportional to the resolution).

2. **For a fixed frame rate constraint, the number of cores required to achieve the lowest total power consumption increases with the video resolution.** For QCIF resolution videos, Fig. 7(d) shows that at frame rates below 15 frames per second (fps) the minimum total power consumption is achieved by 1 core, while above 15 fps the minimum power consumption is achieved by 2 cores. Moreover, using 4 or 8 cores wastes energy because the additional cores consume leakage power, but provide more resources than are necessary to achieve full QoS at QCIF resolution. For CIF resolution videos, Fig. 5 shows that 1 core achieves minimum total power for frame rates below 5 fps, 4 cores achieves minimum power for frame rates between 5 and 25 fps, and 8 cores achieves minimum power for greater than 25 fps.

3. **Higher resolution frames can be partitioned into more slices, enabling more efficient use of available cores (i.e. higher frame rates and lower power consumption).** In the simulation results illustrated in Fig. 7 and Fig. 8, each QCIF resolution frame is partitioned into 4 slices, while in Figures 5 and 6, each CIF resolution frame is partitioned into 8 slices. The improved efficiency achieved by a large number of slices in the CIF resolution video – enabled by improved load balancing – is responsible for the phenomenon described in point 2 above (i.e., the fact that using more cores can actually reduce the total power consumption required to achieve a fixed frame rate).

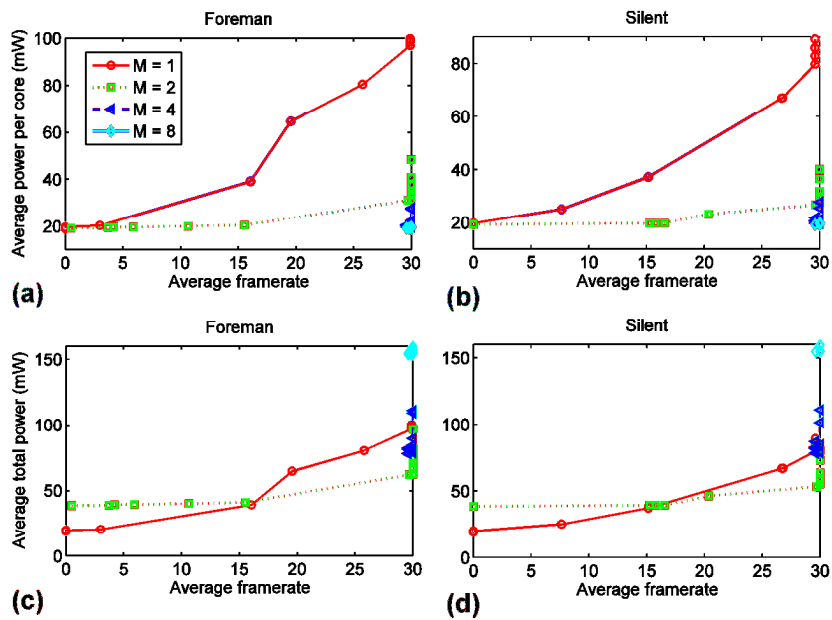

**Fig. 7.** Power consumption versus the average decoded frame rate for QCIF resolution sequences. (a,b) Average power consumption per core versus the average decoded frame rate. (c,d) Average total power consumption versus the average decoded frame rate.

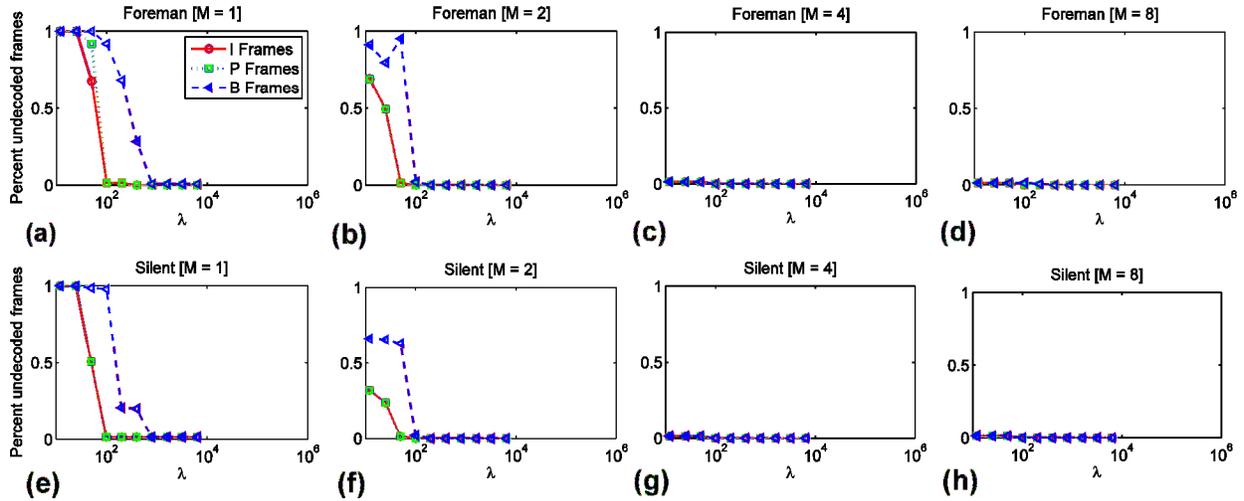

**Fig. 8.** Fractions of I, P, and B frames that miss their display deadline as a function of the parameter $\lambda$ with M = 1, 2, 4, and 8 processors. (a,b,c,d) QCIF resolution Foreman sequence. (e,f,g,h) QCIF resolution Silent sequence.

### 4.5. Experimental comparison

In Fig. 9, we compare our proposed algorithm (with $\lambda = 400$) to the so-called Optimum Minimum-Energy Multicore Scheduling algorithm (OPT-MEMS [3]), and to a modification of our algorithm (with $\lambda = 400$) where we require all processors to operate at the same frequency (i.e. coordinated DVFS). We note that [3] supports both DPM and coordinated DVFS; however, we only compare against the DVFS part to achieve a fair comparison.[9] As in [3], we switch idle processors to the minimum frequency to avoid wasting energy.

OPT-MEMs uses a frame's worst-case execution complexity and its deadline to determine a DVFS schedule that multiplexes between two frequencies in time in order to execute exactly the worst-case number of cycles before the task's deadline. There are four important limitations of OPT-MEMS. First, OPT-MEMS is myopic because it does not consider characteristics and requirements of future tasks (e.g. deadlines, complexities, dependencies) when deciding the DVFS schedule for the current task. Myopic DVFS schedules are known to be suboptimal [20]. Second, OPT-MEMS does not provide a scheduling technique to allocate tasks to processor cores; instead, it assumes that each task is perfectly divisible among an arbitrary number of cores (i.e. it uses a fluid model). This corresponds to the case of perfect load balancing, which can only be achieved in practice if the number of slices per frame is exactly the number of cores, and each slice has exactly the same decoding complexity.[10] Third, OPT-MEMS does not provide a mechanism for scheduling slices belonging to different frames at the same time. This leads to some inefficiency because fully parallelized decoding (which appropriately accounts for frame dependencies) is not possible. Forth, OPT-MEMS uses coordinated DVFS, i.e. it assumes that all processor cores operate at the same frequency. This leads to inefficiency in practice because tasks cannot be perfectly load balanced.

---

[9] Although DPM can be integrated into our proposed solution, we omitted it in this report to simplify the exposition. We leave the integration of DPM into our framework as future research.

As illustrated in Fig. 9(a) and Fig. 9(b), for M = 1 or 2 processors, all algorithms achieve approximately the same frame rates and power consumptions for a given sequence. This is because, even at the highest operating frequency, there are not enough resources to decode all frames. For M = 4 or 8 processors, Fig. 9(a) and Fig. 9(b) show that all algorithms achieve the full frame rate (or very close to the full frame rate); however, Fig. 9(c) and Fig. 9(d) show that the proposed algorithm achieves lower overall power consumption. For M = 4 cores, the proposed algorithm reduces power by approximately 24% for Foreman and 36% for Silent, relative to OPT-MEMS. For M = 8 cores, the proposed algorithm reduces power by approximately 12% for Foreman and 24% for Silent, relative to OPT-MEMS. The improvements are more modest for M = 8 cores because each core runs at a much lower operating frequency than with M = 4 cores, so there is less opportunity to reduce power consumption. It is noteworthy that the change in power consumption between the OPT-MEMS and coordinated DVFS algorithms is largely due to the MDP-based optimization, whereas the change in power consumption between the coordinated DVFS and proposed algorithms is largely due to independent DVFS frequencies for each core (however, the MDP-based optimization and the gains due to independent DVFS frequencies are not completely separable).

---

[10] More precisely, OPT-MEMS defines a speed-up factor $S[j]$ to describe the speed-up achieved (relative to 1 core) when there are $j$ cores available. If a task that takes $w$ cycles on one core is executed in parallel on $j$ cores with a speed-up of $S[j]$, then the task can be executed within $\left\lceil w\,/\,S[j] \right\rceil$ cycles. We assume that $S[j] = 1$ in our comparison, which implies perfect load balancing.

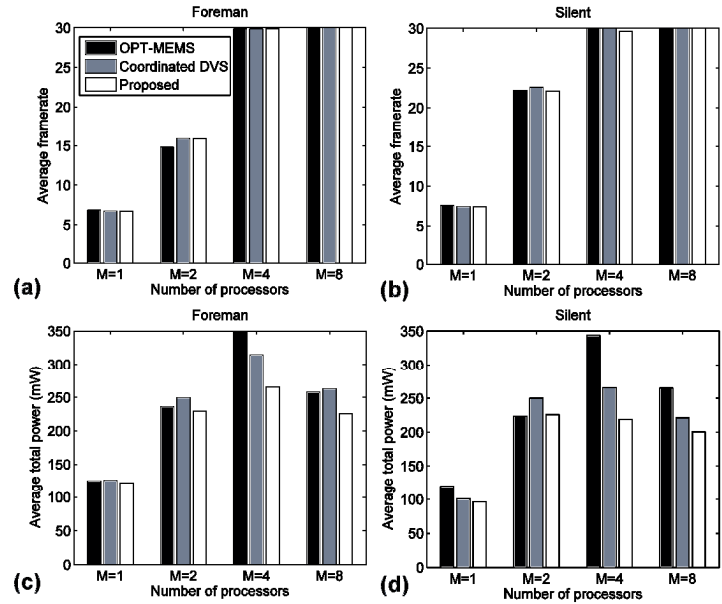

**Fig. 9. Experimental comparisons on M = 1, 2, 4, and 8 processors. (a,b) Average decoded frame rates for Foreman and Silent, respectively. (c,d) Average total power consumption for Foreman and Silent, respectively.**

## 5. CONCLUSION

We propose a Markov decision process based on-line scheduling algorithm for slice-parallel video decoders on multicore systems. Solving for the optimal on-line scheduling and DVFS policy requires complexity that exponentially increases with both the number of processors and the number of frames in a short look-ahead window used by the scheduler. To mitigate this complexity, we proposed a novel two-level scheduler. The first-level scheduler determines scheduling and DVFS policies independently for each frame and the second-level decides the final frame-to-processor and frequency-to-processor mappings at run-time, ensuring that certain system constraints are satisfied. We validated the proposed algorithm in Matlab using accurate video decoder trace statistics generated from a parallelized H.264 decoder that we implemented on a cycle-accurate MPARM simulator. Our experimental results indicate that the proposed algorithm effectively trades-off power consumption and QoS by ensuring that a limited energy-budget is allocated to decoding the most important frames (e.g. I and P frames) before the less important frames (e.g. B frames).

In future work, we plan to integrate the proposed two-level scheduler into the MPARM simulator, first by creating a "hook" between the simulator and Matlab, which will allow us to control the scheduling and DVFS actions at run-time with our Matlab code, and later by actually implementing the two-level scheduler on the master core, which will allow us to measure the impact of the scheduler's overheads on the system's performance. We also plan to integrate DPM into the proposed solution to achieve even lower power consumption.

APPENDIX A: IMPACT OF THE EXPONENTIAL ASSUMPTION

In this appendix, we first describe why we make the exponential assumption. We then provide some analysis to explain the impact of the assumption on the optimal policy. We will refer to Fig. 10 throughout our discussion.

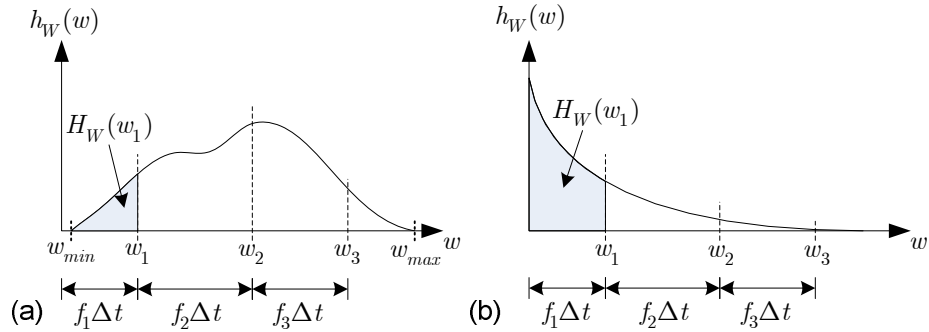

**Fig. 10. (a) General complexity distribution. (b) Exponential complexity distribution.**

### Why model the complexity distribution as an exponential?

Suppose that we have $T$ time slots of duration $\Delta t$ seconds to decode one video slice with random complexity $W \geq 0$ (cycles). Note that the deadline information is embedded in the transition of the current frame sets. Let $h_W(w)$ denote the probability density function (PDF) of $W$ and $H_W(w) = \int_{x=0}^{w} h_W(x)dx$ denote the cumulative density function (CDF) of $W$. In time slot 1, we decode the slice at frequency $f_1$ for $\Delta t$ seconds, executing a total of $w_1 = f_1 \Delta t$ cycles. The probability that the slice is decoded in time slot 1 is $H_W(w_1) = H_W(f_1 \Delta t)$; meanwhile, the probability that the slice is not decoded in time slot 1 is $1 - H_W(w_1) = 1 - H_W(f_1 \Delta t)$. In the special case that $W$ is exponentially distributed with parameter $1/\beta$, i.e. $h_W(w) = \frac{1}{\beta} \exp\left(-\frac{1}{\beta} w\right)$, we have:

$$H_W(w_1) = 1 - \exp\left(-\frac{1}{\beta} w_1\right) \text{ and } 1 - H_W(w_1) = \exp\left(-\frac{1}{\beta} w_1\right).$$

If the slice is not decoded in time slot 1, i.e. $W > w_1$, then, in time slot 2, we decode the slice at a frequency $f_2$ for $\Delta t$ seconds, executing $f_2 \Delta t$ additional cycles. The conditional probability of successfully decoding the slice in time slot 2, given that it was not decoded in time slot 1 (i.e. $W > w_1$), is

$$H_W(w_2 \mid W > w_1) = \frac{\int_{x=w_1}^{w_2} h_W(x) dx}{1 - H_W(w_1)}$$

where $w_2 = f_1 \Delta t + f_2 \Delta t$ denotes the cumulative cycles executed over time slots 1 and 2. Clearly, for a general complexity distribution, the conditional probability of decoding a slice in time slot 2 depends on $w_1$, which is the number of cycles processed in time slot 1. More generally, the conditional probability of decoding a slice in time slot $t$ depends on $w_{t-1}$, which is the cumulative number of cycles processed in time slots 1 through $t-1$.

As shown above, for general complexity distributions, the decoding probabilities for a slice depend on how much that slice was decoded in previous time slots. Consequently, if we use general complexity distributions for the video slices in our MDP framework, then we need to keep track of the number of cumulative cycles processed for every slice in every frame in order to accurately model the slice decoding probability. This requires many additional state variables (one for each slice), making the energy-efficient multicore scheduling problem computationally intractable. However, in the special case that $W$ is exponentially distributed, the conditional probability of decoding a slice in time slot $t$ simplifies to:

$$H_W(w_t \mid W > w_{t-1}) = \frac{\frac{1}{\beta} \int_{x=w_{t-1}}^{w_t} \exp\left(-\frac{1}{\beta} x\right) dx}{\exp\left(-\frac{1}{\beta} w_{t-1}\right)} = 1 - \exp\left(-\frac{1}{\beta}\left(w_t - w_{t-1}\right)\right) = 1 - \exp\left(-\frac{1}{\beta} f_t \Delta t\right),$$

which is independent of $w_{t-1}$ by the memoryless property of the exponential distribution. In this special case, the probability of decoding a slice in any time slot is always the same (conditioned on the operating frequency), so we do not need to maintain any additional state variables. This is why we chose to model the complexity $W$ as an exponential with expected value $E[W] = \beta$, where, in our experiments, we set $\beta$ to the expected value of the true complexity distribution.

**How does modeling the complexity distribution as an exponential impact the final policy?**

As before, assume that we have $T$ time slots of duration $\Delta t$ seconds to decode one video slice with random complexity $W \geq 0$ (cycles). Further assume that the slice decoding complexities are bounded in the interval $\left[w_{min}, w_{max}\right]$ for $0 < w_{min} \leq w_{\max} < \infty$, which is true in practice. Consider the conditional probability of decoding a slice in time slot $t$ given that it was not decoded in time slots 1 through $t-1$ (i.e. $W > w_{t-1}$):

$$H_W(w_t \mid W > w_{t-1}) = \frac{\int_{x=w_{t-1}}^{w_t} h_W(x)dx}{1 - H_W(w_{t-1})}.$$

What happens to $H_W(w_t \mid W > w_{t-1})$ if $w_t \geq w_{max}$?

- Under the true distribution of $W$, if $w_t \geq w_{max}$, then $H_W(w_t \mid W > w_{t-1}) = 1$. In other words, the slice is guaranteed to finish decoding in time slot $t$.

- Under the exponential model of $W$, if $w_t \geq w_{max}$, then $H_W(w_t \mid W > w_{t-1}) = H_W(w_t - w_{t-1}) < 1$ for finite $w_t$. In other words, under the exponential model, there is a finite probability that the slice will not finish decoding in time slot $t$. Consequently, for time slots $t$ near the deadline $T$ (where it is most likely that $w_t > w_{max}$) the optimal policy may[11] be more aggressive than it needs to be because it believes that there is a finite probability of failing to decode the slice (which can possibly incur large costs due to losses of child frames). This aggressive policy may use a higher processor frequency and more power than would be optimal using the true distribution of $W$.

What happens to $H_W(w_t \mid W > w_{t-1})$ if $w_t < w_{min}$?

- Under the true distribution of $W$, if $w_t < w_{min}$, then $H_W(w_t \mid W > w_{t-1}) = 0$. In other words, the slice will not finish decoding in time slot $t$.

- Under the exponential model of $W$, if $0 \leq w_t < w_{min}$, then $H_W(w_t \mid W > w_{t-1}) = H_W(w_t - w_{t-1}) > 0$. In other words, under the exponential model, there is a finite probability that the slice will finish decoding in time slot $t$. Consequently, for time slots $t$ near time slot 1 (where it is most likely that $w_t < w_{min}$) the optimal policy may be more conservative than it needs to be because it believes that there is a finite probability of decoding the slice. This conservative policy may use a lower processor frequency and less power than would be optimal using the true distribution of $W$.

From the above, we can see that using an exponential model of slice decoding complexity may result in an unnecessarily conservative policy in early time slots and an excessively aggressive policy closer to the deadline. Overall, these policies approximately average out in terms of cycles allocated to decoding the slice (relative to the true optimal policy), but end up using more power than necessary (due to the convexity of the power-frequency function).

---

[11] We say that the policy "may" deviate from optimal because, given the discretized set of processor frequencies, it is not always guaranteed to deviate from the optimal solution.

APPENDIX B: COMPUTATIONAL OVERHEADS

There are four components of the proposed algorithm that incur overheads:

1. Offline: Frame-level value iteration at the first-level scheduler (i.e. **Table 3**) with decomposed value iteration update (see Section 3.1.2).

2. Offline: Determining an approximately optimal policy for each frame (i.e. **Table 4**).

3. Online: Policy look-up.

4. Online: Second-level scheduler (see Section 3.2).

The first two components are performed offline, so their overheads will not impact the online system performance. The second two components are performed online, but are light-weight so they require little overhead. For completeness, we discuss the computation overheads of each component below:

**Offline: frame-level value iteration at the first level scheduler**

The proposed frame-level value iteration algorithm requires computing a value function $V^v\left(\mathcal{C}, x^v, r^v\right)$ for each frame $v$ in the group of picture structure $\mathcal{V}^g$. For a single frame $v$, one iteration of the algorithm requires looping through $C$ current frame sets, $l^{\max} + 1$ possible buffer states (a maximum of $l^{\max}$ slices per frame), and two (2) dependency states (i.e. dependencies satisfied or not). For each current frame set $\mathcal{C}$, buffer state $x^v$, and dependency state $r^v$, the algorithm proceeds in $M$ stages, which we refer to as sub-value iterations. The sub-value iteration at processor $M$ requires searching over (a maximum of) $M$ possible previously decoded slices, $F$ processor frequencies and two (2) scheduling actions (i.e., 0 or 1 slice scheduled) and, for each possible combination, requires computing an expectation over two (2) possible departures (i.e., 0 or 1 slice decoded) and a sum over the value functions of frames that are children of frame $v$ (say $U$ children). The sub-value iterations for processors 1 through $M-1$ have the same complexity as the sub-value iteration at processor $M$, except that they do not require computing a sum over the value functions of child frames. Hence, performing $I$ iterations of frame-level value iteration to compute $\left\{V^v\left(\mathcal{C}, x^v, r^v\right) : \forall v \in \mathcal{V}^g\right\}$ requires a number of operations proportional to

$$O\left(2^2 \cdot I \cdot \left|\mathcal{V}^g\right| C \cdot (l^{\max}+1) \cdot F \cdot (2 \cdot M^2 + U)\right)$$

In the experimental results, we consider a four frame group of picture structure (i.e. $\left|\mathcal{V}^g\right| = 4$), twelve current frame sets (i.e. $C = 12$), eight slices per frame (i.e. $l = 8$), four operating frequencies (i.e. $F = 4$), up to eight processors (i.e. $M = 8$), and up to two child frames (i.e. $U = 2$). In our experiments, the frame level value iteration algorithm required at most seventeen iterations to converge (i.e. $I = 17$). Hence, the number of operations to determine the optimal value function is approximately $2^2 \cdot 17 \cdot 4 \cdot 12 \cdot (8+1) \cdot 4 \cdot (2 \cdot 8^2 + 2) = 15275520$.

**Offline: Determining an approximately optimal policy for each frame**

The proposed algorithm for determining an approximately optimal policy for frame $v$ requires computing $\pi^{v,*}\left(\mathcal{C}, x^v, r^v\right)$ given $V_\lambda^{v,*}\left(\mathcal{C}, x^v, r^v\right)$, which was computed above. For a single frame $v$, this requires looping through $C$ current frame sets, $l^{\max} + 1$ possible buffer states, and two (2) dependency states. For each current frame set $\mathcal{C}$, buffer state $x^v$, and dependency state $r^v$, we need to determine $\left(f^{jv,*}, y^{jv,*}\right)$ for all $j \in \{1, \dots, M\}$ (i.e. the optimal frequency and scheduling action at each processor). For processor 1, we find $\left(f^{1v,*}, y^{1v,*}\right)$ as the argument that maximizes the right-hand side of the sub-value iteration at processor 1 in (18) : this requires looping through $F$ processor frequencies and two (2) scheduling actions, and, for each possible combination, requires computing an expectation over two (2) possible departures. For processors $j = 2, \dots, M-1$, we find $\left(f^{jv,*}, y^{jv,*}\right)$ as the argument that maximizes the right-hand side of the sub-value iteration in (17): this requires the same complexity as for processor 1, plus an additional expectation calculation over two (2) possible departures. Finally, for processor $M$, we find $\left(f^{jv,*}, y^{jv,*}\right)$ as the argument that maximizes the right-hand side of the sub-value iteration in (15): this requires the same complexity as for processors $j = 2, \dots, M-1$, plus a sum over the value functions of frames that are children of frame $v$ (say $U$ children). Overall, computing the policy for each frame requires a number of operations proportional to

$$O\left(2 \left|\mathcal{V}^g\right| C \cdot (l^{\max} + 1) \cdot \left(2^2 \cdot M \cdot F + 2 \cdot (M-1) + 2 \cdot F \cdot U\right)\right).$$

Using the parameters of our experiments, the number of operations to determine the policy is approximately $2 \cdot 4 \cdot 12 \cdot (8+1) \cdot \left(4 \cdot 8 \cdot 4 + 2 \cdot 4 \cdot 2 + 2 \cdot (8-1)\right) = 136512$.

**Online: policy look-up**

Online, we use the policy $\pi^{v,*}\left(\mathcal{C}, x^v, r^v\right)$, which was computed above, as a look-up table to determine $\left(\mathbf{f}^{1:M,v,*}, \mathbf{y}^{1:M,v,*}\right) = \pi^{v,*}\left(\mathcal{C}, x^v, r^v\right)$ (i.e. the optimal frequency and scheduling actions for each processor) for all $C$ frames in the current frame set $\mathcal{C}$. Assuming that one table look-up takes $O(1)$, then, in each time slot, we incur an overhead of $O(C)$ to access the policy look-up tables for each frame in the current frame set $\mathcal{C}$.

**Online: Second-level scheduler**

On each of the $M$ processors, the second-level scheduler uses an Earliest Deadline First policy to determine which of the $C$ frames in the current frame set $\mathcal{C}$ get scheduled. This incurs complexity overheads $O(CM)$.

## REFERENCES

[1] H. Aydin and Q. Yang, "Energy-Aware Partitioning for Multiprocessor Real-Time Systems," *Proc. of the 17th International Symposium on Parallel and Distributed Processing* (IPDPS '03), Apr. 2003.

[2] J. M. López, M. García, J. L. Díaz, and D. F. García, "Worst-Case Utilization Bound for EDF Scheduling on Real-Time Multiprocessor Systems," *12th Euromicro Conference on Real-Time Systems (Euromicro-RTS 2000)*, pp. 25 – 33, Jun. 2000.

[3] W. Y. Lee, Y. W. Ko, H. Lee, and H. Kim, "Energy-efficient scheduling of a real-time task on DVFS-enabled multi-cores," *Proc. of the 2009 International Conference on Hybrid Information Technology (ICHIT '09)*, pp. 273-277, 2009.

[4] Y.-H. Wei, C.-Y. Yang, T.-W. Kuo, S.-H. Hung, and Y.-H. Chu, "Energy-efficient real-time scheduling of multimedia tasks on multi-core processors," *Proc. of the 2010 ACM Symposium on Applied Computing (SAC '10)*, pp. 258-262, 2010.

[5] H. Liu, Z. Shao, M. Wang, and P. Chen, "Overhead-Aware System-Level Joint Energy and Performance Optimization for Streaming Applications on Multiprocessor Systems-on-Chip," *Proc. of the 2008 Euromicro Conference on Real-Time Systems (ECRTS '08)*, pp. 92-101, July 2008.

[6] C. E. Leiserson and J. B. Saxe, "Retiming synchronous circuitry," *Algorithmica*, vol. 6, no. 1-6, pp 5–35, 1991.

[7] R. Xu, R. Melhem, and D. Mosse, "Energy-Aware Scheduling for Streaming Applications on Chip Multiprocessors," *Proc. of the 28th IEEE International Real-Time Systems Symposium (RTSS '07)*, pp. 25-38.

[8] R. Xu, "Energy-aware scheduling for streaming applications," Ph.D. Dissertation: http://etd.library.pitt.edu/ETD/available/etd-03082010-201840/unrestricted/XuRuibin20100104.pdf

[9] P. Pillai and K. G. Shin, "Real-time dynamic voltage scaling for low-power embedded operating systems," *SIGOPS Oper. Syst. Rev.*, vol. 35, no. 5, pp. 89-102, Oct. 2001.

[10] D. Zhang, F. Chen, S. Jin, "Global EDF-based online, energy-efficient real-time scheduling in multi-core platform," *2011 IEEE International Conference on Computer Science and Automation Engineering (CSAE)*, vol. 2, pp. 666-670, 10-12 June 2011.

[11] J. Cong and K. Gururaj, "Energy efficient multiprocessor task scheduling under input-dependent variation," *Proc. of the Conference on Design, Automation and Test in Europe (DATE '09)*, pp. 411-416, 2009.

[12] F. Catthoor, E. de Greef, and S. Suytack, *Custom Memory Management Methodology: Exploration of Memory Organisation for Embedded Multimedia System Design.* Kluwer Academic Publishers, Norwell, MA, USA, 1998.

[13] M. Roitzsch, "Slice-balancing H.264 video encoding for improved scalability of multicore decoding," *Proc.of the 7$^{th}$ ACM & IEEE International Conference on Embedded Software*, pp. 269-278, 2007.

[14] R. S. Sutton and A. G. Barto, "Reinforcement learning: an introduction," Cambridge, MA:MIT press, 1998.

[15] L. Benini, D. Bertozzi, A. Bogliolo, F. Menichelli, and M. Olivieri, "MPARM: Exploring the Multi-Processor SoC Design Space with SystemC," *J. VLSI Signal Process. Syst.*, vol. 41, no. 2, pp. 169-182, Sept. 2005.

[16] Real-Time Executive for Multiprocessor Systems (RTEMS): http://www.rtems.org

[17] L. Benini, A. Bogliolo, G. A. Paleologo, G. De Micheli, "Policy optimization for dyamic power management," *IEEE Trans. on Computer-Aided Design of Integrated Circuits and Systems*, vol. 18, no. 6, June 1999.

[18] D. Niyato, S. Chaisiri, L. B. Sung, "Optimal power management for server farm to support green computing," *9$^{th}$ IEEE/ACM International Symposium on Cluster Computing and the Grid*, 2009.

[19] E. B. van der Tol, E. G. Jaspers, R. H. Gelderblom, "Mapping of H.264 decoding on a multiprocessor architecture," *Proc. of the SPIE* (May 2003), pp. 707-718.

[20] E. Akyol and M. van der Schaar, "Complexity Model Based Proactive Dynamic Voltage Scaling for Video Decoding Systems," *IEEE Trans. Multimedia*, vol. 9, no. 7, pp. 1475-1492, Nov. 2007.

[21] W. Yuan, K. Nahrstedt, S. V. Adve, D. L. Jones and R. H. Kravets, "GRACE-1: Cross-layer adaptation for multimedia quality battery energy," *IEEE Trans. Mobile Comput.*, vol. 5, no. 7, pp. 799–815, Jul. 2006.